---------------------------------------------------------------------------
\documentstyle[preprint,eqsecnum,aps]{revtex}

\begin{document}

\title{\bf Scattering theory of superconductive tunneling in quantum
junctions.}

\author{ V. S. Shumeiko $^{a,b}$, E. N.  Bratus'$^{a,b}$, and G. Wendin
$^a$}

\address{
$^a$ Department of Applied Physics, Chalmers University of Technology and
G\"{o}teborg University, S-41296 G\"{o}teborg, Sweden \\
$^b$ B.Verkin Institute for Low Temperature Physics and Engineering,
47 Lenin Ave., 310164 Kharkov, Ukraine}
\maketitle
\draft

\begin{abstract}
We present a consistent theory of superconductive tunneling in single-mode
junctions within a scattering formulation of Bogoliubov-de Gennes quantum
mechanics. Both dc Josephson effect and dc quasiparticle transport in 
voltage biased junctions are considered. Elastic quasiparticle scattering 
by the junction determines equilibrium Josephson current. We discuss the
origin of Andreev bound states in tunnel junctions and their role in 
equilibrium Josephson transport. In contrast, quasiparticle tunneling in 
voltage biased junctions is determined by inelastic scattering.
We derive a general expression for inelastic scattering amplitudes
and calculate the quasiparticle current at all voltages with emphasis on a 
discussion of the properties of subgap tunnel current and the nature 
of subharmonic gap structure. 

\end{abstract}

\section{Introduction.}

The tunnel Hamiltonian model \cite{Bar} has for many years been a main
theoretical tool for investigation of tunneling phenomena in
superconductors \cite{Baro}. However, interpretation of recent experiments
on transmissive tunnel junctions \cite{Fod,Pag,Kle} and  complex
superconductor-semiconductor structures \cite{Tak85,Ova}
require more detailed knowledge of the mechanisms of the
superconductive tunneling than the tunnel model is able to provide.
Particularly informative are experiments on superconducting
quantum point contacts with controlled number of transport modes and
transparency, such as  controllable superconducting break junctions
\cite{Jan} and gate controlled  superconductor-semiconductor devices
\cite{Tak}. Since only a few transport modes with controlled
transparency are involved in the tunnel transport, the experiments
provide precise and detailed information which can be directly
compared with theory.

The first attempts to develop a theory of superconductive
tunneling beyond the tunnel Hamiltonian model \cite{Hab,Za2,Arn1,Asl}
were made in generalization of methods applied to SNS
junctions \cite{Ish,SAB} and superconducting constrictions
\cite{KO,AV} based on Green's function techniques. In these theories, the
junction Green's functions are directly found from Green's function
equations supplemented by special boundary conditions representing the
tunnel barrier, or by matching the superconductor and insulator
Green's functions at the superconductor-insulator boundaries.

In the first works on the Josephson effect in SNS junctions \cite{Ku,BJ}
another way of calculation has been used, based on expansion
over eigenstates of the Bogoliubov-de Gennes (BdG) equation \cite{dG}. A
similar
method has been also applied to SIS tunnel junctions \cite{Svi} and
superconductor-semiconductor junctions \cite{She}. In the absence of
inelastic scattering the method of using the BdG equation
gives the same results as the Green's function technique \cite{Svi}.
One might then expect that the Josephson effects in
superconducting junctions can be explained on a rather simple quantum
mechanical level. Following this idea, the quantum mechanical approach 
has been successfully applied to calculation of the dc Josephson current
in different kinds of mesoscopic weak links
\cite{Bee1,Bag,Bee2,Hurd,Smslt} and tunnel junctions \cite{Fur,BeeH,Sh}.
The first application of the method to voltage biased 
junctions was done by Blonder, Tinkham, and Klapwijk who considered
quasiparticle tunneling in SIN junctions as a scattering problem in BdG
quantum mechanics \cite{BTK}. Later on, the quantum mechanical approach
has been found to be helpful in investigations of more complex
phenomena of quasiparticle transport and ac Josephson effect in voltage
biased SNS junctions \cite{KBT}, mesoscopic SIS tunnel junctions \cite{BShW}
and mesoscopic constrictions \cite{Ave}.

The quantum mechanical approach based on the BdG equation is adequate for
describing
the physical situation in mesoscopic junctions, where inelastic
scattering effects are weak and coherent electron dynamics
is of main importance. Moreover, due to the effect of quantization of
transverse electron modes in mesoscopic junctions \cite{Bee1,Wee},
1D models for the current transport through the junction may be
appropriate.

In this paper we present a consistent quantum mechanical theory of
superconductive tunneling in a one-mode quantum constriction,  Fig.1. We
consider the dc Josephson effect and also dc quasiparticle tunneling in
voltage
biased junctions. In the latter case we focus attention on a
detailed calculation  of the subharmonic gap structure (SGS) of the
tunnel current \cite{BShW}.

Following the Landauer approach \cite{Lnr}, we
consider superconducting electrodes as equilibrium reservoirs which emit
quasiparticles into the constriction. Scattering by the junction goes
into two channels: (i) the
normal channel with outgoing quasiparticles remaining  in the same
branch of the quasiparticle spectrum,
and (ii) the Andreev channel where the quasiparticles change
branch due to electron-hole conversions.  The current in
such a picture results from the imbalance of currents carried by
scattering states originating from the left and the right reservoirs,
the magnitude of the current being proportional to the transmission
coefficient $D$ of the tunnel barrier.

The imbalance of currents in superconducting junctions can be created in
two ways: by establishing a difference of the
phases of the order parameters in the left and right electrodes, or
by applying a voltage bias. The basic fact concerning the flow of
equilibrium
current in the presence of a phase difference,
established by Furusaki and Tsukada \cite{Fur}, is that a  bulk
supercurrent, when approaching the tunnel interface, is
transformed into current flowing through superconducting bound
states which appear at the tunnel
interface in the presence of this phase difference \cite{Kup} and which
provide transmission of the Cooper pairs through the tunnel barrier.
The  balance among currents of different scattering states is not violated,
although the scattering amplitudes depend strongly on the phase difference.

Application of a voltage bias gives rise to more far reaching
consequences than just imbalance of the elastic scattering
modes: the scattering states themselves are
modified in a non-trivial way. This follows from the
fact that the scattering amplitudes, being phase-dependent in equilibrium,
become time-dependent in accordance
with the Josephson relation \cite{Jos}, $d\phi/dt=2eV$,  when voltage
is applied. Thus, in the presence of a dc voltage the superconducting
junction behaves as an effective nonstationary scatterer whose
transmissivity oscillates. This property of superconducting
junctions gives rise to ac Josephson effect; however, it is also
significant for dc quasiparticle transport because the quasiparticle
transmission through such a scatterer is {\em inelastic}.

The physical mechanism of inelastic quasiparticle transmission through
voltage biased superconducting junctions has been
first considered in SNS junctions \cite{KBT} where it has been explained in
terms of multiple Andreev
reflections (MAR):  the normal quasiparticles confined
between superconducting walls are permanently accelerated by the static
electric field due to sequential electron-hole conversions at the NS
interfaces, similarly to acceleration of the electrons in
an ordinary potential well by a time dependent electric field. Similar
arguments can be extended to the tunnel junctions \cite{Arn2}; however,
in tunnel junctions the scattering theory approach is more
appropriate because of the quantum nature of quasiparticle
transmission through the atomic-size tunnel barrier. This introduces a
side-band spectrum of scattered waves where side-band energies are
shifted with respect to the energy of the
incident wave by integer number of quanta of the scatterer frequency
\cite{BShW}. Such an approach is familiar in the theory of quantum
scattering
by oscillating potential barriers  in normal tunnel junctions (see e.g.,
Refs.  \cite{Sto,Bag1} and references therein).

The tunneling through all the inelastic channels (normal and Andreev)
constitutes a complete picture of superconductive tunneling in biased
Josephson junctions - the
incoherent part of side band currents corresponding to the dc quasiparticle
current and the side band interference currents
corresponding to the ac Josephson current. An important aspect of this
picture is that the Andreev bound states are involved in the current
transport together with the extended side band states, giving a
multiparticle character to the superconductive tunneling in the subgap
voltage region. This multiparticle origin of the subgap tunnel current
was first pointed out by Schrieffer and Wilkins \cite{SW}.

The structure of the paper is the following. After formulation of the
problem and discussion of the quasiclassical approximation in Sec. II,
we  consider the problem of elastic scattering in Sec. III as a starting
point for
construction of inelastic scattering states in biased junctions. The
solution of the elastic scattering problem allows us to calculate dc
Josephson
current, which is done for completeness in Sec. IV.
In Sec. V we construct inelastic
scattering states and derive a continued-fraction representation for the
scattering amplitudes. Sec. VI is devoted to derivation of
the nonequilibrium current. In Sec. VII we discuss the origin of the excess
tunnel current in the large bias
limit. In Sec. VIII we present a general analysis of the subgap tunnel
current.
Finally, the SGS is analyzed in Sec. IX.

\section{Formulation of model.}

We consider a superconducting quantum constriction with adiabatic
geometry \cite{Gla}: the cross section varies smoothly with
the coordinate $x$ on the  scale of the Fermi electron wave length,  
$1/p_F$, and the size of cross section is comparable with the Fermi electron
wave length, Fig. 1. The length $L$ of the constriction is assumed to be 
smaller than the superconducting coherence length $\xi_0$, 

\begin{equation}
1/p_F\ll L \ll \xi_0.
\end{equation}
The Hamiltonian of the constriction is assumed to have the form:
\begin{equation}
\hat H=\bigg[\frac{(\hat{\vec p}-\sigma_ze\vec A(\vec r,t))^2}{2m}+
U(\vec r)-\mu\bigg]\sigma_z+
\left[V(x)+e\varphi(\vec r,t)\right] \sigma_z + \hat \Delta(\vec r,t),
\label{Hamiltonian}
\end{equation}
where $U(\vec r)$ is the potential confining electrons within the constriction,
$V(x)$ is the potential of the tunnel barrier, 
$\vec A(\vec r,t)$ and $\varphi(\vec r,t)$ are electromagnetic
potentials, and $\hat \Delta(\vec r,t)$ is the off-diagonal superconducting 
order parameter given by the matrix:
\begin{equation}
\hat \Delta=
\left(
\begin{array}{cc}
0 & \Delta e^{i\chi/2}\\
\Delta e^{-i\chi/2} & 0
\end{array}\right).
\label{Delta}
\end{equation}
We assume the junction to be symmetric. The choice of the units corresponds to
$c=\hbar=1$.

It is convenient to eliminate the
phase of the superconducting order parameter $\chi(\vec r,t)$ in Eq. (\ref{Delta})
by means of a gauge transformation:
\begin{equation}
e^{i\sigma_z\chi/2} \hat H e^{-i\sigma_z\chi/2} \rightarrow \hat H, 
\label{gauge}
\end{equation}
which allows us to introduce a
gauge invariant superfluid momentum, $\vec p_s=\nabla\chi/2-e\vec A$ and 
an electric potential $\Phi=\dot{\chi} /2+e\varphi$. 

There are different scales of change of potentials in Eq. (\ref{Hamiltonian}): 
one is an atomic scale over which the confining potential $U(r_\perp)$ and the 
potential of the tunnel barrier $V(x)$ change. Other scales
are related to change of superconducting order parameter, 
electromagnetic field penetration lengths and length of the contact: all these
lengths  are large in comparison with the atomic length. It is convenient to
separate these two scales, introducing quasiclassical wave functions 
\cite{Ba2} which vary slowly on an atomic scale, and including rapidly 
varying potentials in a boundary condition for quasiclassical wave functions. 
To this end we assume the solution $\Psi(\vec r,t)$ of the Bogoliubov-de 
Gennes equation \cite{dG}
\begin{equation}
i\dot\Psi(t)=\hat H\Psi(t),
\label{BdG0}
\end{equation}
with the Hamiltonian of Eq. (\ref{Hamiltonian}), 
to have the quasiclassical form
\begin{equation}
\Psi(\vec r,t)=\sum_{\beta}\psi_\perp(\vec r_\perp,x){1\over\sqrt{v}}
e^{i\beta\int pdx}\psi^\beta(x,t),
\label{quasiclassic}
\end{equation}
where $\psi_\perp$ is the normalized wave function of the quantized transverse
electron motion with the energy $E_\perp$,
\[({\hat{\vec p}_\perp \over 2m}+ V(\vec r_\perp,x))\psi_\perp=
E_\perp(x)\psi_\perp,
\;\;\;\psi_\perp(r_\perp=\infty,x)=0,\]
and $p$ is the longitudinal momentum of the quasiclassical electron,
$p(x)=\sqrt{2m(\mu-E_\perp(x))}$; $\beta=\pm$ indicates the direction of 
the electron motion. We assume here that the constriction has only one 
transport mode; an extension to the case of several unmixed modes consists of
additional summation  over all transport modes in the equation for the current.
The coefficients $\psi^\beta$ in Eq. (\ref{quasiclassic}) describe  
the wave functions slowly varying in the $x$ direction and satisfy the
reduced BdG equation
\begin{equation}
i\dot\psi^\beta_{L,R}=(\beta v\hat p\sigma_z + \Phi_{L,R}\sigma_z +
vp_{sL,R} + \Delta\sigma_x)\psi^\beta_{L,R}
\label{BdG}
\end{equation}
in the the left (L) and the right (R) electrodes; $v=p/m$.
The potentials $\vec p_s$ and $\Phi$ describe the distributions of
electromagnetic
field and supercurrent in the electrodes. In the point contact geometry
these quantities are small due to the effect of spreading out of current
\cite{KO,KOSh}, and we will omit them, $\vec p_s=\Phi=0$. For the same
reason, deviation of spatial distribution of the module of the order 
parameter $\Delta$ from constant magnitude is small in the point contacts, 
and we will neglect it, $\Delta=const$.

The functions $\psi^\beta_{L,R}$ are matched at the constriction 
by the boundary condition \cite{Sh} (see also Appendix A): 
\begin{equation}
\left(\begin{array}{c}
\psi_L^-\\
\psi_R^+\\
\end{array}\right) 
=\hat V
\left(\begin{array}{c}
\psi_L^+\\
\psi_R^-
\end{array}\right)\;\;\mbox{at}\;\; x=0\;,
\label{match}
\end{equation}
with a matching matrix $\hat V$ 
\begin{equation}
\hat V=
\left(
\begin{array}{cc}
r & de^{i\sigma_z\phi/2}\\
de^{-i\sigma_z\phi/2} & r
\end{array}\right).
\label{V}
\end{equation}
$d$ and $r$ are the normal electron transmission and reflection
amplitudes due to the barrier, and $\phi$ is a 
gauge invariant difference of the superconducting phases of right
and left electrodes: 
$\phi=\chi_R(0)-\chi_L(0).$ 
The matching matrix in Eq. (\ref{V}) satisfies the unitarity condition
\begin{equation}
\hat V\hat V^\dagger=1.
\label{unitary}
\end{equation}
The boundary condition in Eqs. (\ref{match}), (\ref{V}) is analogous to the 
boundary condition used in the quasiclassical Green's functions techniques 
(see e.g., \cite{Za2,KuL}). This is the simplest equation for coupling of 
superconducting electrodes, while retaining the main features of the Josephson
effect, except for effects of the resonant tunneling \cite{BeeH,KuD,WSh}.

\section{Elastic scattering }

In the absence of time  dependence in the phase difference at the junction, 
$\dot\phi=0$, Eqs. (\ref{BdG}) and (\ref{match}) describe elastic 
scattering of quasiparticles. The scattering states are to be constructed 
using stationary solutions of Eq.(\ref{BdG}) which correspond to elementary 
propagating waves with energy $|E|>\Delta$:
\begin{mathletters}
\label{propagating}
\begin{equation}
\psi^{\beta\alpha}_E=e^{-iEt+i\beta\alpha(\xi/v)x}u^{\delta}_E,\;\;
\end{equation}
\begin{equation}
u^{\delta}_E={1\over \sqrt{2\cosh\gamma}}\left(
\begin{array}{c}
e^{\delta\gamma/2}\\
\sigma e^{-\delta\gamma/2}
\end{array}\right),
\end{equation}
\end{mathletters}
where
\begin{equation}
\xi=\sqrt{E^2-\Delta^2},\;\;
e^\gamma={|E|+\xi \over \Delta},
\sigma=\mbox{sign} E, \;\;\alpha=\pm,\;\;\delta=\alpha\sigma.
\end{equation}
The vector function $u_E$ is normalized, $(u\,,\,u)=1$, the brackets meaning
scalar product in electron-hole space.
In Eq. (\ref{propagating}) there are four elementary waves which 
correspond to the same energy,  as illustrated in Fig. 2, which are labeled
by quantum numbers $\beta$ (direction of the Fermi electron momentum) and 
$\alpha= sign(|p|-p_F)$ (electron or hole-like branch of the quasiparticle
spectrum).
The direction of propagation of each elementary wave is determined by the sign
of the probability current.
The probability current density $j_p$, defined by the conservation law
(continuity equation)
$\partial|\psi|^2/\partial t+\partial j_p/\partial x=0$ for the BdG equation 
Eq. (\ref{BdG}), has the form
$j_p=(\psi,\sigma_z\psi)$. For the elementary waves 
in Eq. (\ref{propagating}) one obtains the explicit result 
$j_p=\beta\delta\tanh\gamma$. 
According to this formula, the relation $\delta=\beta$ is met for for the
waves propagating from left to right, $\delta=-\beta$ for the waves
propagating from right to left. Therefore the incoming waves
from the left (L) and right (R) have the form
\begin{equation}
L:\;\; e^{i\sigma(\xi/v)x}u^\beta_E, \;\;\;
R:\;\; e^{-i\sigma(\xi/v)x}u^{-\beta}_E,
\label{in}
\end{equation}
while the outgoing waves have the form
\begin{equation}
L:\;\; e^{-i\sigma(\xi/v)x}u^{-\beta}_E, \;\;\;
R:\;\; e^{i\sigma(\xi/v)x}u^\beta_E.\\
\label{out}
\end{equation}
Correspondingly, the incoming quasiparticle can be
scattered into four outgoing states: two forward scattering states and two
back scattering states. One of the reflected waves belongs to the same
(electron- or hole-like) branch of the quasiparticle spectrum as the
incoming wave and constitutes the normal scattering channel, while the other
reflected wave changes spectrum branch and constitutes the Andreev
channel. In a similar way transmitted waves constitute normal and 
Andreev channels. The structure of the scattering states then becomes
\begin{mathletters}
\label{scattering}
\begin{equation}
\left(
\begin{array}{l}
\psi^-_L\\ \psi^+_R
\end{array}\right)=
\left( \begin{array}{l}
\delta_{j,1}\\ \delta_{j,2}
\end{array}\right)
e^{i\sigma(\xi/v)x}u^-_E +
\left( \begin{array}{l}
a\\b
\end{array}\right)_j
e^{-i\sigma(\xi/v)x}u^+_E,
\end{equation}
\begin{equation}
\left( \begin{array}{l}
\psi^+_L\\ \psi^-_R
\end{array}\right)=
\left( \begin{array}{l}
\delta_{j,3}\\ \delta_{j,4}
\end{array}\right)
e^{i\sigma(\xi/v)x}u^+_E +
\left( \begin{array}{l}
c\\f
\end{array}\right)_j
e^{-i\sigma(\xi/v)x}u^-_E
\end{equation}
\end{mathletters}
(for brevity we have omitted the time dependent factors $e^{-iEt}$).
In Eqs. (\ref{scattering}) the index $j=1(2)$  corresponds to a hole-like
quasiparticle coming from the left(right), while
index $j=3(4)$ corresponds to an electron-like quasiparticle  coming from 
the left (right).
According to the structure of the matching matrix Eq. (\ref{V}), there is
the following symmetry between the scattering states $j=1$ and $j=2$:
\begin{equation}
\left( \begin{array}{c}
a\\b
\end{array}\right)_2 (\phi)=
\left( \begin{array}{c}
b\\a
\end{array}\right)_1(-\phi), \;\;
\left( \begin{array}{c}
c\\f
\end{array}\right)_2 (\phi)=
\left( \begin{array}{c}
f\\c
\end{array}\right)_1 (-\phi).
\label{symmetry21}
\end{equation}
The analogous symmetry exists also for the scattering states $j=3,4$.
Using the unitarity of the matching matrix Eq. (\ref{V}) one can find the
following relation between scattering states $j=3$ and $j=1$:
\begin{equation}
\left( \begin{array}{c}
a\\b
\end{array}\right)_3 (\gamma,r,d)=
\left( \begin{array}{c}
c\\f
\end{array}\right)_1 (-\gamma,r^\ast,d^\ast),\;\;\;
\displaystyle
\left( \begin{array}{c}
c\\f
\end{array}\right)_3 (\gamma, r,d)=
\left( \begin{array}{c}
a\\b
\end{array}\right)_1 (-\gamma, r^\ast,d^\ast).
\label{symmetry31}
\end{equation}
These symmetry relations allow us to find all the scattering amplitudes if 
one of the scattering states is known.

Let us find the explicit scattering amplitudes for the scattering state $j=1$.
After substituting Eqs. (\ref{scattering}) into Eq. (\ref{match}) it is 
convenient to split the resulting equation, using the orthogonality condition,
$(u^+,\sigma_z u^-)=0$, into two independent equations for the normal
scattering amplitudes $c,f$ and for the Andreev scattering amplitudes $a,b$,
\begin{mathletters}
\label{normal}
\begin{equation}
(u^-,\sigma_z u^-)
\left( \begin{array}{l}
1\\0
\end{array} \right)=
(u^-,\sigma_z \hat V u^-)
\left( \begin{array}{l}
c\\f
\end{array} \right)_1,
\end{equation}
\begin{equation}
(u^+,\sigma_z u^+)
\left( \begin{array}{l}
a\\b
\end{array} \right)_1=
(u^+,\sigma_z \hat V u^-)
\left( \begin{array}{l}
c\\f
\end{array} \right)_1.
\end{equation}
\end{mathletters}
Calculating the scalar products in Eqs. (\ref{normal}) we find
explicit expression for the Andreev amplitudes in terms of the normal ones,
\begin{equation}
\left(
\begin{array}{l}a\\b
\end{array}
\right)_1={id\sin (\phi/2) \over \sinh \gamma}
\left(
\begin{array}{r}f\\-c
\end{array}
\right)_1.
\label{ab}
\end{equation}
The solution of the first equation in Eq. (\ref{normal}) is given by
\begin{equation}
c_1={r\sinh^2\gamma \over Z},\;\;\;
f_1=-{d\sinh\gamma\sinh(\gamma+i\phi/2) \over Z},
\label{cf}
\end{equation}
where
\begin{equation}
Z=-{d\over d^\ast}
\Big(R\sinh^2\gamma+D\sinh(\gamma+i\phi/2)\sinh(\gamma-i\phi/2)\Big),
\label{Det}
\end{equation}
$D=|d|^2$ is the normal electron transmission coefficient of the tunnel 
junction, and $R=|r|^2=1-D$ is  the normal electron reflection coefficient.
It follows from Eqs. (\ref{ab}), (\ref{cf}) 
that if there is no phase difference across the junction, $\phi=0$, the 
Andreev scattering channel is closed: $a=b=0$. It is worth mentioning that
the Andreev reflection is
also absent if the normal transparency of the junction is equal to zero,
$D=0$. If, on the other hand, the junction is completely transparent for
normal electrons, $D=1$, there is no Andreev forward scattering, $b=c=0$.

In the presence of a phase difference at the junction the quasiparticle
scattering is accompanied by the appearance of superconducting bound states
\cite{Kup}. One can establish the existence of bound states by investigating 
the poles of the scattering amplitudes, Eq. (\ref{cf}), at imaginary $\gamma$ 
corresponding to energies lying inside the gap $|E|<\Delta$.
Assuming $\gamma\rightarrow i\gamma$ in Eq. (\ref{Det}),
we have the dispersion equation $Z(i\gamma)=0$, or
\begin{equation}
\sin^2\gamma=D\sin^2\phi/2.
\label{dispersion}
\end{equation}
The bound states correspond to a positive magnitude of $\sin\gamma$: 
$ \Delta\sin\gamma=\mbox {Im}\;\xi >0$. This condition  selects two roots:
\begin{equation}
\gamma=\gamma_0=\arccos(\sqrt{D}\sin\phi/2),\;\; \gamma=\pi-\gamma_0,
\end{equation}
or
\begin{equation}
E(\phi)=\pm\Delta\sqrt{1-D\sin^2\phi/2}.
\label{spectrum}
\end{equation}
The wave functions of the bound states are to be constructed from
elementary solutions of Eq. (\ref{BdG}) with $|E|<\Delta$ which decay at
$x=\pm\infty$,
\begin{mathletters}
\label{bound1}
\begin{equation}
\varphi^\beta_{E,R}=e^{-iEt-\zeta x/v}
u_E^{\nu},
\end{equation}
\begin{equation}
\varphi^\beta_{E,L}=e^{-iEt+\zeta x/v}
u_E^{-\nu},
\end{equation}
\end{mathletters}
where
\begin{equation}
u_E^{\nu}=
{1\over\sqrt{2}}
\left(
\begin{array}{c}
e^{i\nu\gamma/2}\\
\sigma e^{-i\nu\gamma/2}
\end{array}\right),\;\;\;
e^{i\gamma}={|E|+i\zeta\over\Delta},\;\;\zeta=\sqrt{\Delta^2-E^2},\;\;
\nu=\beta\sigma.
\label{bound2}
\end{equation}
The bound state ansatz has a form similar to the outgoing part of the
scattering states Eq. (\ref{scattering})
with the coefficients satisfying the homogeneous equations of Eq.
(\ref{normal}). These coefficients are:
\begin{mathletters}
\label{bound abcf}
\begin{equation}
f=-{d\sin(\gamma+\phi/2)\over r\sin\gamma}\;c,
\end{equation}
\begin{equation}
\left(
\begin{array}{l}a\\b
\end{array}
\right)={d\sin (\phi/2) \over \sin \gamma}
\left(
\begin{array}{l}f\\-c
\end{array}
\right),
\end{equation}
\end{mathletters}
with $\gamma$ given by Eq. (\ref{dispersion}). We note that the bound
state spectrum is nondegenerate.
The coefficient $c$ in Eqs. (\ref{bound abcf}) is obtained from the
normalization condition for the bound state wave function,
$$
\int d^2r_\perp \int_{-\infty}^{\infty} dx |\Psi|^2=
{1\over\zeta}(|a|^2+|b|^2+|c|^2+|f|^2)=1,
$$
which yields
\begin{equation}
|c|^2=\Delta\sin\gamma\left( 1+
{D\over R}{\sin^2(\gamma+\phi/2)\over \sin^2\gamma} \right)^{-1}.
\label{bound c}
\end{equation}

What is the origin of the bound states in a tunnel junction? 
According to Eq. (\ref{normal}) one can regard these states as resulting 
from hybridization of the bound states in the short ballistic constriction 
\cite{Bee1} due to the normal electron reflection by the barrier
(cf. effect of impurities in the SNS junction \cite{Bag,Bee2}). 
Let us consider a smooth constriction with the length exceeding the
coherence length, $L\gg\xi_0$. In such a constriction the supercurrent
density and the superfluid momentum are related by the local equation, 
$J_s(x)=(e/m)N_sp_s(x)$, and they are
both enhanced in the neck of the constriction due to current
concentration (for simplicity we neglect the effect of suppression of 
the superfluid electron density $N_s$ by the supercurrent). The local 
quasiparticle spectrum in the
presence of supercurrent has an additional contribution  $\pm v_Fp_s(x)$
\cite{dG}, which gives rise
to a shift of the local energy gap, Fig.3. The spatial bending of the gap
edges forms the potential wells at $E<0$($E>0$) for quasiparticles with 
electron velocities directed along (opposite) the current. 
The bound states in these potential wells are similar to the Andreev bound 
states in the SNS junctions \cite{Andr}, the difference being that here
the bound states are caused by the spatial nonhomogeneity of the {\em phase}
of the order parameter, while the original Andreev states are
caused by the spatial nonhomogeneity of the modulus of the order
parameter. With decreasing length of the constriction the number of the
bound states in the well decreases. The short Josephson constriction
corresponds to an infinitely narrow and deep $\delta$-potential well which
contains only one Andreev level \cite{Bee1}.

\section{dc Josephson current}

A convenient expression for the tunnel current results from statistical
averaging of the current operator written in the Nambu representation 
\cite{Schr}: 
\begin{equation}
I(x,t)={e\over 2m}\left\{(\hat p- \hat p')\int d^2r_\perp
\left[ \delta (\vec r -\vec r')  
-Tr\;\langle\hat\Psi(\vec r,t)\;
\hat\Psi^\dagger (\vec r',t)\rangle\right]\right\}_{\vec r=\vec r'},
\label{current1}
\end{equation}
where $\hat\Psi$ is a two-component field operator:
\begin{equation}
\hat\Psi(\vec r,t)=
\left(\begin{array}{c}
\hat\psi_\uparrow(\vec r,t)\\
\hat\psi^\dagger_\downarrow(\vec r,t)
\end{array}\right),
\end{equation}
and $Tr$ is a trace in electron-hole space.
The angular brackets in  Eq. (\ref{current1}) denote a thermal average
of the one-particle density matrix of the superconductor \cite{Gal}.
In equilibrium this matrix has the form
\begin{equation}
\langle\hat\Psi(\vec r)\; \hat\Psi^\dagger (\vec r')\rangle
=\sum_\lambda
\Psi_\lambda(\vec r)n_F(-E_\lambda)\Psi^\dagger_\lambda (\vec r'),
\label{density matrix}
\end{equation}
where $ \Psi_\lambda(\vec r)$ are the eigenstates of the stationary BdG
equation  Eq. (\ref{BdG0}) with the quantum numbers $\lambda$.
We note that the definition of Fermi distribution function $n_F$ here 
corresponds to
the distribution of holes in the normal metal: in the ground state all energy
levels above the Fermi level ($E>0$) are occupied, while energy levels
below the Fermi level ($E<0$) are empty (see also the discussion in the 
next section).
In the quasiclassical approximation, Eq. (\ref{quasiclassic}), the average
tunnel current calculated at the middle of the junction has the form 
\begin{equation}
I=-e\sum_\lambda n_F(-E_\lambda)\sum_\beta \beta|\psi_{\lambda}^{\beta}(0)|^2.
\label{current2}
\end{equation}
The current in Eq. (\ref{current2}) can be calculated either at the left or 
the right side of the junction because,
due to the unitarity of the matching matrix $\hat V$ in Eq. (\ref{unitary}),
the equality
\begin{equation}
|\psi_{L}^+|^2-
|\psi_{L}^-|^2=
|\psi_{R}^+|^2-
|\psi_{R}^-|^2
\label{continuity}
\end{equation}
holds for the each eigenstate.
The current in Eq. (\ref{current2}) consists of contributions from both 
the scattering states and the bound states
\begin{equation}
I=-\int_{|E|>\Delta}{dE|E|\over 2\pi\xi}n_F(-E)\sum_j I_j(E)
-\sum_{|E|<\Delta} n_F(-E)I_{bound}(E),
\label{current3}
\end{equation}
$$
I(E)=e\sum_{\beta}\beta|\psi^{\beta}(E)|^2.
$$
When calculating the contribution from the scattering states, it is convenient 
to consider the transmitted current of each scattering mode:
\begin{equation}
I_j(E)=\left\{
\begin{array}{lr}
e(|b_j|^2-|f_j|^2) & j=1,3\\
e(|c_j|^2-|a_j|^2) & j=2,4
\end{array}
\right. .
\label{Ejcurrent}
\end{equation}
The symmetry relations Eqs. (\ref{symmetry21}), (\ref{symmetry31}) yield
\begin{equation}
I_1(E)= I_4(E),\;\;
I_2(E)=I_3(E),\;\;
I_2(E)=-I_1(E).
\label{jsymmetry}
\end{equation}
Thus the currents of all the scattering states with a given energy cancel each
other in equilibrium \cite{note1}. 
Substitution of equations Eq. (\ref{bound abcf}), (\ref{bound c}) into Eq.
(\ref{current3}), taken e.g. at the right electrode, yields for the current
of the bound state
\begin{equation}
I_{bound}(E)= e(|b|^2-|f|^2)=-{e\Delta^2\over 2E}D\sin\phi.
\label{bound current}
\end{equation}
A useful formula for the current of the single bound state, which allows 
direct evaluation of the current from the bound state spectrum, is given by
equation:
\begin{equation}
I(E)=2e{dE(\phi)\over d\phi}
\label{dE/dphi},
\end{equation}
where $E(\phi)$ is the bound state energy band, Eq.  (\ref{spectrum}). This
formula is derived in Appendix B.
Taking into account Eqs. (\ref{bound current}), (\ref{current3}), the total 
current has the form \cite{Hab,Za2,Arn1,AB}:
\begin{equation}
I={e\Delta D\sin\phi\over 2\sqrt{1-D\sin^2\phi/2}}
\tanh{\Delta\sqrt{1-D\sin^2\phi/2}\over 2T}.
\label{AB}
\end{equation}

Thus, the dc Josephson current in tunnel junctions is carried only by
the bound states, which is similar to the situation found in the 
other kinds of short weak links \cite{BeeH,Bee1,Bag,Bee2,Smslt}.
It follows from Eqs. (\ref{current3}), (\ref{bound current}) that 
the nonvanishing total current results from the imbalance of the bound state 
currents due to a difference in the equilibrium population numbers.
Creation of a nonequilibrium population gives rise to a possibility to 
control the Josephson transport \cite{Sh,WSh,Smslt}.

\section{Inelastic scattering}

Now we proceed to a discussion of inelastic scattering in voltage biased
junctions. According to our assumption $\Phi=0$, explained in Sec. II, 
the applied voltage drop $V$ is confined to the constriction; we also 
neglect, in order not to complicate the problem, a small time dependent
voltage induced across the junction by the ac Josephson current (self-coupling
effect \cite{Wer}). This implies the following dependence  on time of the
phase difference:
\begin{equation}
\phi=\phi_0+2eVt.
\label{phase}
\end{equation}
The appearance of factors with periodic time dependence in the boundary 
condition, Eqs.  (\ref{match}),(\ref{V}), gives rise to a more complex
structure of the scattering states than in Eq. (\ref{scattering}).
In order to satisfy the boundary condition, the outgoing 
part of the scattering states in
Eq.(\ref{scattering}) is to be constructed from the eigenstates  
of equation Eq. (\ref{BdG}) with different energies $E_n=E-neV$ shifted
with respect to the energy $E$ of the incoming wave with an integer 
$-\infty<n<\infty$ (sideband structure)
\begin{mathletters}
\label{nscattering}
\begin{equation}
\pmatrix{\psi^-_L\cr \psi^+_R\cr}(0)=
\pmatrix{\delta_{j,1}\cr \delta_{j,2}\cr}u^-_Ee^{-iEt} + \sum_{n}
\pmatrix{a\cr b\cr}_{j,n} u^+_ {E_n}e^{-iE_nt},
\end{equation}
\begin{equation}
{\psi^+_L\choose \psi^-_R}(0)=
{\delta_{j,3}\choose \delta_{j,4}}u^+_Ee^{-iEt} + \sum_{n}
{c\choose f}_{j,n} u^-_{E_n}e^{-iE_nt}.
\end{equation}
\end{mathletters}
For brevity we use the notation $u_n=u_{E_n}$. While the incoming state is 
itinerant, the outgoing 
states can be either itinerant [Eq. (\ref{propagating}) if $|E_n|>\Delta$]
or bound [Eq. (\ref{bound2}) if $|E_n|<\Delta$]. It is convenient to combine
both the equations for the functions $u_n$ in a single analytical form:
\begin{equation}
u_n^\pm=
\displaystyle {1\over\sqrt{2\cosh\Gamma_n}}
\left(
\begin{array}{c}
e^{\pm\gamma_n/2}\\
\sigma_n e^{\mp\gamma_n/2}
\end{array}
\right)
\end{equation}
\begin{equation}
e^{\gamma_n}=\displaystyle {|E_n|+\xi_n\over\Delta},\;\;
\Gamma_n=\mbox{Re}\gamma_n,\;\;
\xi_n=
\left\{
\begin{array}{lr}
\sqrt{E^2_n-\Delta^2}, &  |E_n|>\Delta\cr
i\sigma_n\sqrt{\Delta^2-E^2_n}, &  |E_n|<\Delta. 
\end{array}\right.
\end{equation}
To find the scattering amplitudes in Eq. (\ref{nscattering}) we consider
the boundary condition Eq. (\ref{match}). It is important to mention, that
this boundary condition was derived neglecting the energy dispersion of the
normal electron scattering amplitudes $d$ and $r$, which means that now this
assumption should be valid for the entire interval of relevant energies
$E_n$.  Let us first discuss $j=1$ (hole-like quasiparticle coming from 
the left): 
$$
\pmatrix{1\cr 0\cr}u_n^-\delta_{n,0}+\pmatrix{a\cr b\cr}_{1,n}u_n^+=
r \pmatrix{c\cr f\cr}_{1,n}  u_n^- 
+\displaystyle{d\over2}\pmatrix{1+\sigma_z &0\cr 0&1-\sigma_z\cr}
\pmatrix{f\cr c\cr} _{1,n-1}u_{n-1}^-
$$
\begin{equation}
+\displaystyle{d\over2}\pmatrix{1-\sigma_z &0\cr 0&1+\sigma_z\cr}
\pmatrix{f\cr c\cr} _{1,n+1}u_{n+1}^-.
\label{matching1}
\end{equation}

It is convenient to separate the equations for normal and Andreev 
scattering amplitudes in Eq. (\ref{matching1}) again using a
procedure similar to Eq. (\ref{normal}). 
The equation for the normal scattering amplitudes then becomes 
\begin{equation}
rc_{1,n}+(d/2)(V^-_{nn+1}f_{1,n+1}+V^+_{nn-1}f_{1,n-1})=\delta_{n0}\\
\label{nnormal}
\end{equation}
$$
rf_{1,n}+(d/2)(V^+_{nn+1}c_{1,n+1}+V^-_{nn-1}c_{1,n-1})=0,
$$
where the coefficients
\begin{equation}
V^\pm_{nm}=\displaystyle {(u^{-\ast}_n,\:\sigma_z\pm 1\:\:u^-_m)\over 
(u^{-\ast}_n,\:\sigma_z\:u^-_n)}\\
\end{equation}
have the explicit form
\begin{mathletters}
\label{Vn}
\begin{equation}
V^+_{nm}=-\displaystyle{e^{-(\gamma_n+\gamma_m)/2} \over \sinh\gamma_n}
\sqrt{\cosh\Gamma_n \over \cosh\Gamma_m},\\
\end{equation}
\begin{equation}
V^-_{nm}=\sigma_n\sigma_m\displaystyle{e^{(\gamma_n+\gamma_m)/2} \over 
\sinh\gamma_n} \sqrt{\cosh\Gamma_n \over \cosh\Gamma_m}.
\end{equation}
\end{mathletters}
The equation for the Andreev scattering amplitudes reads
\begin{equation}
a_{1,n}=(d/2)(U^-_{nn+1}f_{1,n+1}+U^+_{nn-1}f_{1,n-1})
\label{nandreev}
\end{equation}
$$
b_{1,n}=(d/2)(U^+_{nn+1}c_{1,n+1}+U^-_{nn-1}c_{1,n-1})
$$
where the coefficients  are defined as
\begin{equation}
U^\pm_{nm}=\displaystyle{(u^{+\ast}_n,\:\sigma_z\pm 1\:\:u^-_m)\over 
(u^{+\ast}_n,\:\sigma_z\:u^+_n)},\\
\end{equation}
and have the explicit forms
\begin{mathletters}
\label{Un}
\begin{equation}
U^+_{nm}=\displaystyle{e^{(\gamma_n-\gamma_m)/2} \over \sinh\gamma_n}
\sqrt{\cosh\Gamma_n \over \cosh\Gamma_m},\\
\end{equation}
\begin{equation}
U^-_{nm}=-\sigma_n\sigma_m\displaystyle{e^{-(\gamma_n-\gamma_m)/2} \over 
\sinh\gamma_n} \sqrt{\cosh\Gamma_n \over \cosh\Gamma_m}.
\end{equation}
\end{mathletters}
As can be seen from  Eqs. (\ref{nnormal}),(\ref{nandreev}), the inelastic
scattering possesses a specific asymmetry: the forward scattered waves have
odd side band indices and backward scattered waves have 
even side band indices,  as illustrated in Fig. 4. Correspondingly, bound 
states with 
odd or even side band indices are induced either in the right or in the left
electrode. We note that the scattering to any side band consists of both
normal and Andreev components. 

It is instructive to compare the superconducting scattering 
diagram in Fig. 4 with the scattering diagram of normal junctions.
In the normal limit $\Delta=0$,  all the
Andreev amplitudes in Eq. (\ref{nandreev}) vanish 
[$U^\pm_n=0$ in  Eq. (\ref{Un})] 
and equations Eq. (\ref{nnormal}) split due to 
$V^+=0$ in Eq. (\ref{Vn}), which yields
$f_{n}=c_{n-1}=0$ for all $n\neq 1$. Thus, the side band diagram in Fig. 4 reduces to the
elementary fragment shown in Fig. 5a. This fragment corresponds to the
scattering of a {\em true hole}, meaning a  
particle with spectrum $E_h=-(p^2/2m-\mu)$ according to the BdG equation Eqs.
(\ref{Hamiltonian}),(\ref{BdG0}). In the ground state, $T=0$, these holes
fill all positive energy states $E>0$ while the negative energy
states are empty. For the electrons, the corresponding diagram is sketched in 
Fig. 5b. In this diagram the chemical potentials in both electrodes are equal, 
while the energies of the incident and transmitted states are shifted by $eV$.
This difference from the conventional diagram of normal electron
tunneling in Fig. 5c (where the chemical potentials in the electrodes are
shifted relative to each other, while the scattering is elastic)
appears after separating out the superconducting phase in Eq. (\ref{gauge}); 
the conventional picture with shifted chemical potential can be restored 
by means of the gauge transformation of the normal
electron wave function $\psi\rightarrow \exp(-ieVt)\psi$. 

When 
superconductivity is switched on, $\Delta\neq 0$, the incoming quasiparticle
consists of both electron and hole components, and therefore the scattering
diagram is a combination of the diagrams in Figs. 5a,b. One has also to take 
into account electron-hole conversions, which lead to the appearance of both 
electron and hole components in the upper as well as lower transmitted 
state. Continuation of this process creates the whole superconducting 
scattering diagram in Fig. 4.

From a mathematical point of view, Eqs. (\ref{nnormal}),(\ref{nandreev}) 
for the scattering amplitudes are second order
difference equations which can not be solved exactly, except in  
some particular cases, e.g.
a fully transparent constriction ($r=0$) where equation  Eq. (\ref{nnormal})
reduces to a binary relation \cite{Ave}. In the general case, it is only 
possible 
to find asymptotical solutions using some small parameter. In the present 
case of a tunnel junction, there is a natural small parameter -
the transparency of the tunnel barrier: $D\ll 1$. However, a straightforward
perturbation expansion with respect to this parameter gives rise to 
divergences, 
similar to difficulties of the multiparticle tunneling theory (MPT) 
\cite{SW,W,Has}. In order to formulate an improved perturbation procedure, 
it is convenient to rewrite Eqs. (\ref{nnormal}),(\ref{nandreev}) 
in terms of the parameter $\lambda=D/4R$, the true small parameter of
the theory, as will be seen later. To this end we introduce new scattering 
amplitudes 
\begin{equation}
c_{1,\pm 2k}={\lambda^k\over r}c_{\pm 2k},\;\;\;
f_{1,\pm (2k+1)}={\lambda^k d^\ast\over 2R}f_{\pm (2k+1)},\;\;\;
a_{1,\pm 2k}=\lambda^k a_{\pm 2k},
\label{new1}
\end{equation}
$$
b_{1,\pm(2k+1)}={\lambda^k d\over 2r}b_{\pm(2k+1)},
$$
which satisfy the equations:
\begin{mathletters}
\label{equation}
\begin{equation}
c_n+\lambda V^-_{nn+1}f_{n+1}+V^+_{nn-1}f_{n-1}=0\\
\end{equation}
$$
f_n-\lambda V^+_{nn+1}c_{n+1}-V^-_{nn-1}c_{n-1}=0,
$$
\begin{equation}
a_n=\lambda U^-_{nn+1}f_{n+1}+U^+_{nn-1}f_{n-1}\\
\end{equation}
$$
b_n=\lambda U^+_{nn+1}c_{n+1}+U^-_{nn-1}c_{n-1},
$$
for $n>0$. For $n<0$ one has to make the change 
$V^\pm_{nm} \rightarrow V^\mp_{-n-m} \;\;
U^\pm_{nm} \rightarrow U^\mp _{-n-m},\;\; (n,m>0)$ in the above equations.
The equation for $n=0$ then reads
\begin{equation}
c_0+\lambda(V^-_{01}f_1+V^+_{0-1}f_{-1})=1
\end{equation}
\end{mathletters}

Let us now turn to the second scattering case in Eq. (\ref{nscattering}), $j=2$
(hole-like quasiparticle incoming from the right). 
According to the symmetry relations of Eq. (\ref{symmetry21}) the scattering
amplitudes $j=2$ differ
from the scattering amplitudes $j=1$ by $\phi\rightarrow -\phi$, which in
the present time-dependent case means that $n\pm 1\rightarrow n\mp 1$. 
Taking into account
this symmetry and also the property of scattering amplitudes in Eq.
(\ref{symmetry21}), we introduce new scattering amplitudes
\begin{equation}
f_{2,\pm 2k}={\lambda^k\over r}\overline c_{\pm 2k},\;\;\;
c_{2,\pm (2k+1)}={\lambda^k d^\ast\over 2R}\overline f_{\pm (2k+1)},\;\;\;
b_{2,\pm 2k}=\lambda^k \overline a_{\pm 2k},
\label{new2}
\end{equation}
\[
a_{2,\pm(2k+1)}={\lambda^k d\over 2r}\overline b_{\pm(2k+1)},
\]
which satisfy the following equations (for $n>0$):
\begin{mathletters}
\label{equation'}
\begin{equation}
\overline c_n+\lambda V^+_{nn+1}\overline
f_{n+1}+V^-_{nn-1}\overline
f_{n-1}=0\\
\end{equation}
$$
\overline f_n-\lambda V^-_{nn+1}\overline c_{n+1}-V^+_{nn-1}\overline
c_{n-1}=0,
$$
\begin{equation}
\overline a_n=\lambda U^+_{nn+1}\overline f_{n+1}+U^-_{nn-1}
\overline f_{n-1}\\
\end{equation}
$$
\overline b_n=\lambda U^-_{nn+1}\overline c_{n+1}+U^+_{nn-1}
\overline c_{n-1},
$$
\begin{equation}
\overline c_0+\lambda(V^+_{01}\overline f_1+V^-_{0-1}\overline f_{-1})=1.
\end{equation}
\end{mathletters}
Equations (\ref{equation'}) differ from equations Eq. (\ref{equation})
by
\begin{equation}
V^\pm \rightarrow V^\mp,\;\;
U^\pm \rightarrow U^\mp.
\end{equation}

In the case of  electron-like quasiparticles incoming
from the left, $j=3$, the symmetry of Eq. (\ref{symmetry31}) involves 
transformation 
$\gamma\rightarrow -\gamma$, which means transformation of the
coefficients
$V^\pm \rightarrow -\sigma_n\sigma_m V^\mp,\;\;
U^\pm \rightarrow -\sigma_n\sigma_m U^\mp$ in Eqs. (\ref{nnormal}),
(\ref{nandreev}).
This allows us to relate the scattering amplitudes of this case   
to the solutions of Eqs. (\ref{equation'}):
\begin{equation}
a_{3,\pm 2k}={\lambda^k\over r^\ast}\sigma_{\pm 2k}
\overline c_{\pm 2k},\;\;\;
b_{3,\pm (2k+1)}=-{\lambda^k d\over 2R}\sigma_{\pm(2k+1)}
\overline f_{\pm (2k+1)},\;\;\;
c_{3,\pm 2k}=\lambda^k \sigma_{\pm 2k}
\overline a_{\pm 2k},
\label{new3}
\end{equation}
$$
f_{3,\pm(2k+1)}=-{\lambda^k d^\ast\over 2r^\ast}\sigma_{\pm(2k+1)}
\overline b_{\pm(2k+1)}.
$$
In a similar way the scattering amplitudes of electron-like
quasiparticles incoming from the right, $j=4$, are related to the
solutions of Eqs. (\ref{equation}),
\begin{equation}
b_{\pm 2k}^{(4)}={\lambda^k\over r^\ast}\sigma_{\pm 2k}
c_{\pm 2k},\;\;\;
a_{\pm (2k+1)}^{(4)}=-{\lambda^k d\over 2R}\sigma_{\pm (2k+1)}
f_{\pm (2k+1)},\;\;\;
f_{\pm 2k}^{(4)}=\lambda^k \sigma_{\pm 2k} a_{\pm 2k},
\label{new4}
\end{equation}
$$
c_{\pm(2k+1)}^{(4)}=-{\lambda^k d^\ast\over 2r^\ast}
\sigma_{\pm(2k+1)} b_{\pm(2k+1)}.
$$

According to a symmetry of the coefficients in  Eqs. (\ref{equation}),
(\ref{equation'}),
\begin{equation}
V^\pm_{nm}(-E)=V^{\pm\ast}_{-n-m}(E),\;\;
U^\pm_{nm}(-E)=U^{\pm\ast}_{-n-m}(E),
\end{equation}
all scattering amplitudes with positive and negative incoming energies are
related through 
\begin{equation}
a_n(-E)=\overline a^\ast_{-n}(E),
\label{Esymmetry}
\end{equation}
and similarly for the other amplitudes.

Let us now present a formal solution of Eq. (\ref{equation}) for $n>0$ on the 
following form \cite{FTT}:
\begin{equation}
f_{2k+1}=(-1)^k\prod^{2k+1}_{l=0}S_l c_0,
\label{f}
\end{equation}
where the quantities $S_l$ are defined as
\begin{equation}
S_{2k}=-{c_{2k}\over f_{2k-1}}, \;\;\;
S_{2k+1}={f_{2k+1}\over c_{2k}},
\label{defS}
\end{equation}
and satisfy the recurrence relations
\begin{equation}
S_{2k}={V^+_{2k,2k-1}\over 1+\lambda V^-_{2k,2k+1}S_{2k+1}},\;\;\;
S_{2k+1}={V^-_{2k+1,2k}\over 1+\lambda V^+_{2k+1,2k+2}S_{2k+2}}.
\label{S}
\end{equation}
The quantity $c_0$ in Eq. (\ref{fn}) is given by
\begin{equation}
c_0={1\over 1+\lambda(V^-_{01}S_1+V^+_{0-1}S_{-1})}.
\label{c0}
\end{equation}
It is convenient to express the functions $S_n$ in Eq. (\ref{S}) through the
relation $S_{n}=V^\pm_{n,n+1}/Z_n$, were the denominators $Z_n\;(n\neq 0)$ 
satisfy the recurrence
\begin{equation}
Z_n=1+\lambda {
a^\pm_n a^\pm_{n+1} \over Z_{n+1}},\;\;\;
a^\pm_n={e^{\pm\gamma_n}\over \sinh\gamma_n},
\label{Zn}
\end{equation}
($\pm$ corresponds to even/odd $n$), and to define $Z_0$ as
the denominator of $c_0$, Eq. (\ref{c0}):
\begin{equation}
Z_0=1+\lambda {a_0^+a_1^+\over Z_1}+
\lambda {a_0^-a_{-1}^-\over Z_{-1}}.
\label{Z0}
\end{equation}

Using the above notation, one is able to express the
coefficients of the normal forward scattering, $|f_n|^2$, on the form
\begin{equation}
|f_n|^2={e^{\gamma_0} \over \cosh\gamma_0|Z_0|^2}e^{\Gamma_n}\cosh\Gamma_n
\prod_{l=1}^{n}{1 \over |Z_l\sinh\gamma_l|^2},
\label{fn}
\end{equation}
provided $a^-\neq 0$. 
The equation for the coefficients of the normal backward scattering, $|c_n|^2$,
differs from Eq. (\ref{fn}) by $e^{\Gamma_n}\rightarrow e^{-\Gamma_n}$. 
The relation between the amplitudes of the
Andreev and normal forward scattering in Eq. (\ref{equation}), taking into
account Eqs.(\ref{defS}), (\ref{S}), (\ref{Zn}), has the form
\begin{equation}
b_n=-e^{-\gamma_n}\Big(1-\lambda {2e^{-\gamma_{n+1}} \over
\sinh\gamma_{n+1}Z_{n+1} } \Big) f_{n}.
\label{bn}
\end{equation}

In a similar way, one can express the solution of Eq.
(\ref{equation'}) for $n>0$ on the form 
\begin{equation}
|\bar f_n|^2={\displaystyle e^{-\gamma_0} \over 
\displaystyle\cosh\gamma_0|\bar Z_0|^2}
e^{-\Gamma_n}\cosh\Gamma_n\displaystyle
\prod_{l=1}^{n}{\displaystyle 1 \over 
\displaystyle|\bar Z_l\sinh\gamma_l|^2},\;\;\;
\bar b_n=-e^{\displaystyle \gamma_n}\Big(1+\lambda {\displaystyle 
2e^{\gamma_{n+1}} \over
\displaystyle\sinh\gamma_{n+1}\bar Z_{n+1} } \Big) \bar f_{n},
\label{fn'}
\end{equation}
where 
\begin{equation}
\displaystyle
\bar Z_n=1+\lambda {
a^\mp_n a^\mp_{n+1} \over \bar Z_{n+1}},\;\;
\displaystyle
\bar Z_0=1+\lambda {a_0^-a_1^-\over\bar Z_1}+
\lambda {a_0^+a_{-1}^+\over\bar Z_{-1}}.	
\label{Zn'}
\end{equation}
We note that Eqs. (\ref{fn'}), (\ref{Zn'}) differ from Eqs. (\ref{fn}),
(\ref{bn}) by $\gamma_n \rightarrow -\gamma_n$ everywhere.

Equations for the scattering amplitudes with negative side band
indices, $n<0$, can be derived in a similar way, and the result differs from
the above equations for positive side band indices Eqs.
(\ref{Zn})-(\ref{Zn'}) by the substitution
\begin{equation}
\gamma_n\rightarrow-\gamma_{-|n|},\;\;n\neq 0,
\end{equation}
introduced everywhere except in $Z_0$ and  $\bar {Z}_0$.

\section{Quasiparticle current}

In the nonstationary problem under consideration, the density matrix
determining the current Eq. (\ref{current1}) is time dependent, and its 
dynamic evolution can be described by an equation similar to Eq.
(\ref{density matrix}),
\begin{equation}
\langle\hat\Psi(\vec r,t)\; \hat\Psi^\dagger (\vec r',t)\rangle
=\sum_\lambda
\Psi_\lambda(\vec r,t)f_\lambda\Psi^\dagger_\lambda (\vec r',t),
\label{tdensity matrix}
\end{equation}
$\Psi_\lambda$ are now solutions of the time dependent problem,
Eq. (\ref{BdG0}),
with initial conditions corresponding to the eigenstates of the
initial Hamiltonian with the eigenvalues $\lambda$, and 
occupation numbers $f_\lambda$ of these initial states. We consider inelastic
scattering states, Eqs. (\ref{quasiclassic}), (\ref{nscattering}),
as the propagators $\Psi_\lambda(t)$ in Eq. (\ref{tdensity matrix}) with
$\lambda$ corresponding to the complete set of the incoming states
$\lambda=(E,j)$; according to the assumption about local equilibrium within
the electrodes, the incoming states possess the Fermi distribution of
occupation numbers, $f_{Ej}=n_F(-E)$. Thus the current Eq. (\ref{current1})
takes the form:
\begin{equation}
I(t)=-e\int_{|E|>\Delta} {dE\,|E|\over 2\pi\xi}
n_F(-E)\sum^{\infty}_{N=-\infty}
e^{iNeVt} 
\sum ^{\infty}_{n=-\infty}
\sum_{j\beta}\beta (\psi^{\beta}_j(E,n),\,\psi^{\beta}_{j}(E,N+n)).
\label{current2'}
\end{equation}
The current in Eq. (\ref{current2'}) consists of a time independent part,
$N=0$, which is formed by incoherent
contributions of all the side bands (the quasiparticle current), and of a 
time dependent part, $N\neq
0$ which results from interference among the different side bands (ac
Josephson current). The  difference between the side band indices $N$ is
an even number since the side band index is either even or odd depending on
the electrode; therefore the time-dependent current oscillates with the 
Josephson frequency $\omega=2eV$. 

In this paper we will concentrate on an analysis of the time-independent 
quasiparticle current. Similarly to Eq. (\ref{Ejcurrent}) we will calculate
the current using the transmitted states,
\begin{equation}
I={1\over 2}{e\over\pi}\int_{|E|>\Delta} {dE\,|E|\over \xi} n_F(-E)
\sum^{\infty}_{n=-\infty}
\bigg[\sum_{j=1,3}(|f_{jn}|^2-|b_{jn}|^2)
+\sum_{j=2,4}(|a_{jn}|^2-|c_{jn}|^2)\bigg].
\label{current3'}
\end{equation}
Using the scattering amplitudes introduced in the previous section through
Eqs. (\ref{new1}), (\ref{new2}), (\ref{new3}) and (\ref{new4}) we express  
the current in Eq. (\ref{current3'}) in the following form:
\begin{equation}
I={e\over \pi}\int_{|E|>\Delta} {dE\,|E|\over \xi} n_F(-E)
\sum_{odd}
(K_n-\overline K_n),
\label{current4'}
\end{equation}
where
\begin{equation}
K_n=
\lambda^{|n|}
(R^{-1}|f_{n}|^2- |b_{n}|^2),\;\;
\overline K_n=
\lambda^{|n|}
(R^{-1}|\overline f_{n}|^2 -|\overline b_{n}|^2)
=K_n({-\gamma}).
\label{Kn}
\end{equation}
The factor of two appears in Eq. (\ref{current4'}) due to equality of 
currents $I_1$ and $I_4$ and of currents $I_2$ and $I_3$ in
Eq. (\ref{jsymmetry}), equalities which hold also in the nonstationary case. However,
there is no balance between currents of these two pairs any more.
The symmetry of Eq. (\ref{Esymmetry}) allows us to reduce the interval of 
integration in Eq. (\ref{current4'}) to the semiaxis $E>0$, 
\begin{equation}
I={e\over \pi}\int_{\Delta}^{\infty} {dE\,E\over \xi} \tanh{E\over 2T}
\sum_{odd}
(K_n-\overline K_n).
\label{current5'}
\end{equation}
The side band currents $K_n$ in Eq. (\ref{Kn}) are proportional to the powers
of the small parameter $\lambda$, $K_n\sim\lambda^n$. Therefore 
Eqs. (\ref{current5'}), (\ref{Kn}) present a perturbative expansion of the
current, convenient for analysis in the limit of low barrier
transparency.
The following sections are devoted to such an analysis of the structure of the
current in Eq. (\ref{current5'}).

\section{Excess current at large bias}

To make some observations useful for analysis of the subgap current, it is
instructive first to discuss the simpler case of large bias
$eV\gg\Delta$, which is well studied in literature \cite{AV,Za2,BTK,Arn1}. 
Simultaneously we will derive the explicit analytical expression for the
current in this limit valid in a whole range of the junction transparency
$0<D<1$. 
The asymptotic expansion of the current with respect to the small parameter 
$\Delta/eV$ has the form
\cite{AV}: 
\begin{equation}
I={e^2DV\over\pi}+I_{exc}(D)+O\left({\Delta\over eV}\right),
\label{largeI}
\end{equation}
where the first term is the tunnel current of the normal junction and the
second term is a voltage-independent excess current which represents the
leading superconducting correction. 

A main simplification in this case is that the side band currents $K_n$ and
$\overline K_n$, $|n|>1$ diminish when the bias voltage increases. This
follows from an estimate of the transmission amplitudes in Eqs.
(\ref{fn})-(\ref{fn'}), which contain products of factors 
$|\sinh\gamma_k|^{-2}$ which are small at large voltages,
$|\sinh\gamma_k|^{-2}\sim (\Delta/eV)^2$, because of the large interval of
involved energies, $E\sim eV$. Furthermore, inspection of the amplitudes
$f_{-1}$ and $\bar f_1$ shows that they are also small due to the factors
$e^{-\gamma_0-\gamma_1}$; therefore the non-vanishing 
part of the current Eq. (\ref{current5'}) in the limit
$eV\gg\Delta$ becomes
\begin{equation}
I={e\over \pi}\int_{\Delta}^{\infty} {dE\,E\over \xi} \tanh{E\over 2T}
(K_1-\overline K_{-1}).
\label{current5''}
\end{equation}
The essential fragments of the scattering diagram in the large bias limit
are shown in Fig. 6. 

The structure of the current in Eq. (\ref{current5''}) is essentially
determined by the presence of a gap in the spectrum of the side band $n=1$;
this causes different analytical forms of the current $K_1$ in the
regions $|E|<\Delta$ and $|E|>\Delta$. We note that the spectrum of the
side band $n=-1$ possesses no gap: $E_{-1}>\Delta$ for $E>\Delta$.
According to this we divide the integral in Eq.
(\ref{current5''}) into three parts:
\[I=I_{<}+I_{\Delta}+I_{>}.\]
The first part corresponds to the current of the states 
in the side band $n=1$, which lie below the gap, $E_1<-\Delta$.
The second part corresponds to the current of the states of the
same side band lying within the gap, $-\Delta<E_1<\Delta$. 
The third part combines contributions from the remaining states of the
side band $n=1$, $\Delta<E_1$, and from the all states of the side band $n=-1$.
Making use of the approximations 
\begin{equation}
|Z_0|^2=
|\overline Z_0|^2\approx
|1+2\lambda(E+\xi)/\xi|^2,\;\;\;
|Z_{-1}|^2= |\overline Z_1|^2=
|Z_2|^2= |\overline Z_{-2}|^2\approx 1,
\end{equation}
$$
|Z_1|^2=
|\overline Z_{-1}|^2=
|Z_{-2}|^2=
|\overline Z_{2}|^2\approx 1/R^2,
$$
it is possible to express the integral $I_<$ on the form
(we restrict ourselves to the limit $T=0$):
\begin{equation}
I_{<}= {e\over\pi}\int_{\Delta}^{eV-\Delta}
dE{E\over\xi}K_1=
{8e\lambda\over\pi R}\int_{\Delta}^{eV/2}
dE{E\over\xi Z_0^2}
-{2e\lambda\over\pi R}\int_{\Delta}^{\infty}
dE{(E-\xi)\over\xi Z_0^2}\left(
1-4\lambda R{E\over\xi} \right)
\label{I_1'}
\end{equation}
where the limit of integration in the last term is extended to 
infinity since the main contribution to this integral comes from the energies 
$E\sim\Delta\ll eV$.
Separating out the normal junction current we may express
Eq. (\ref{I_1'}) in the following form 
\begin{equation}
I_{<}= {e^2DV\over\pi}-
{8e\lambda^2\over\pi }\int_{\Delta}^{\infty}{dE\over Z_0^2\xi}
\left[ 4\lambda R{E\Delta^2\over\xi^2}
+(2R+1) {E(E-\xi)\over\xi}
+{1\over 4\lambda R}(E-\xi)
\right].
\label{I_1''}
\end{equation}
We note that this current is always smaller than the normal current.
It is convenient to express the integral $I_\Delta$ as
\begin{equation}
I_{\Delta}={e\over\pi}\int_{eV-\Delta}^{eV+\Delta}
dE{E\over\xi}K_1=
{16e\lambda^2\over\pi }\int_0^{\Delta}
dE{\Delta^2\over|\xi Z_0|^2},
\label{I_2'}
\end{equation}
obtained by using the relations between the functions $Z_n$ 
[which result from their definition in Eq. (\ref{Zn})]
\begin{equation}
Z_n(E+eV)=\overline Z_{n-1}(E),\;\;
Z_0(E+eV)Z_1(E+eV)=\overline Z_0(E)\overline Z_{-1}(E).
\label{shift}
\end{equation}
Inspection of the equation for $I_>$, 
\begin{equation}
I_{>}={e\over\pi}\int_{eV+\Delta}^{\infty}
dE{E\over\xi}K_1-
{e\over\pi}\int_{\Delta}^{\infty}
dE{E\over\xi}\overline K_{-1},
\label{I_3}
\end{equation}
shows that both integrals diverge at 
the upper limit $E=\infty$, which means that the states lying far from the
Fermi level formally contribute to the current, while the
quasiclassical approximation of Eq. (\ref{quasiclassic}) assumes, that all
relevant states lie close to the Fermi level. To get rid of this formal 
divergence one commonly shifts by $eV$ the variable in the first
integral in  Eq. (\ref{I_3}). Using again the relations (\ref{shift}) we 
may express this integral on the form
\[
{4e\lambda\over\pi R}\int_{\Delta}^{\infty}
dE{E\over\xi Z_0Z_{-1}}+
{8e\lambda^2\over\pi }\int_{\Delta}^{\infty}
dE{E(E-\xi)\over\xi^2 Z_0^2},
\]
where the first term has the same analytical form but the opposite sign
compared to the divergent term in the second integral in Eq. (\ref{I_3}),
\[
-{4e\lambda\over\pi R}\int_{\Delta}^{\infty}
dE{E\over\xi Z_0Z_{-1}}+
{2e\lambda\over\pi R}\int_{\Delta}^{\infty}
dE{(E-\xi)\over\xi Z_0^2}.
\]
After elimination of the divergent terms, the integral in Eq. (\ref{I_3}) 
takes the form
\begin{equation}
I_{>}=
{2e\lambda\over\pi R}\int_{\Delta}^{\infty}
dE{(E-\xi)\over\xi Z_0^2}\left(
1+4\lambda R{E\over\xi}
\right).
\label{I_3'}
\end{equation}
The currents in Eq. (\ref{I_3'}) and Eq. (\ref{I_2'}) are positive and
overcompensate the missing part of the current in Eq.
(\ref{I_1''}). Collecting Eqs. (\ref{I_1''}),  (\ref{I_2'}) and
(\ref{I_3'}) we find after some algebra the following explicit equation for the
excess current in Eq. (\ref{largeI}),
\begin{equation}
I_{exc}={16e\Delta\lambda^2R\over\pi }
\left[
1-{D^2\over 2(1+R)\sqrt{R}}\ln{1+\sqrt{R}\over 1-\sqrt{R}}
\right],
\label{excess}
\end{equation}
which is valid in the whole interval of junction transparency $0<D\leq 1$.
Asymptotics of this expression coincide with the results presented in
literature \cite{Za2,BTK} both in the limit of fully transparent ($D=1$)
constrictions, $I_{exc}=8e\Delta /3\pi,$ and in the limit of 
low-transparency ($D\ll 1$) tunnel junctions, $I_{exc}=e\Delta D^2/\pi$.

The above calculation reveals an important difference 
between the structure of the current in normal and superconducting
junctions. In normal junctions, the current, e.g. in the
right electrode,  see Fig. 5c, results from 
scattering states lying above the local chemical potential, $E>\mu-eV$, 
while contribution from the energy interval $E<\mu-eV$ is equal to zero due to
mutual cancellation of currents of the scattering states incident from the
left and from the right (in Fig. 5a the current carrying energy region 
corresponds to negative energies $E_h<0$). Thus the total current 
coincides with the current 
of real excitations emitted from the contact, which is consistent with
the nonequilibrium origin of the current in the voltage biased junctions.
In superconducting junctions, only "across-the-gap" current $I_{<}$ is clearly
related to the real excitations emitted at the right side of the junction
where the current is calculated (Fig. 6a) - the dissipative character of the
currents $I_{\Delta}$ and $I_{>}$ is not obvious. However, one
should take into account the creation of real excitations at the left side
of the junction via back scattering into the side band $n=2$ (Fig. 6b,c). 
Although
the current of this side band exists only at the left side of the
scattering diagram it should have an effect at the right side 
due to continuity of the current at the
interface, Eq. (\ref{continuity}), and therefore it should be distributed 
among the states of the side band $n=1$. As our calculations show, this 
"kick" current partially flows through the Andreev bound states, involving
the current $I_{\Delta}$, Fig. 6b, which
convert this current into a supercurrent outside the junction. 
It is also partially distributed among the scattering states with positive 
energies, current $I_{>}$, Fig. 6c, in the form of imbalanced ground
state currents.

\section{Subgap current}

In this section we turn to a discussion of the tunnel current in the subgap
region $eV<2\Delta$. A basic property of the subgap current 
is the presence of temperature independent structures on I-V
characteristics - the subharmonic gap structure (SGS). 
SGS in tunnel junctions was discovered in experiments by Taylor and
Burstein \cite{Bur}, and the first theoretical explanation was given by
Schrieffer and Wilkins \cite{SW} in terms of multiparticle tunneling
(MPT). Recently SGS has been observed in 
a number of experiments on transmissive tunnel junctions \cite{Fod,Pag,Kle}. 
Although there is a possibility to attribute SGS in planar junctions to
normal shorts, the observation of SGS in superconducting  controllable break
junctions \cite{Jan} provided convincing confirmation of the existence of 
SGS in the true tunnel regime.

The existence of SGS in tunnel current can be established within the MPT theory
by means of rather simple
perturbative arguments \cite{SW,W,Has,Wo}. On the basis of the 
tunnel Hamiltonian model, assuming a small perturbative coupling between 
electrodes, one can calculate 
the probability of tunneling in $n$-th order of perturbation 
theory. Such a probability is proportional to a product of filling factors 
of the initial and the final states: $n_F(E)[1-n_F(E-neV)]$. At zero
temperature this factor is equal to zero
outside the interval $\Delta<E<neV-\Delta$, which selects the quasiparticle 
transitions across the gap, i.e. the processes of 
creation of real excitations relevant for the tunnel current. Such a
restriction places the threshold of the $n$-th
order current at $eV=2\Delta/n$, and a sequence of current onsets of 
magnitudes $\sim D^n$ at the voltages $eV=2\Delta/n$ forms the SGS of the 
tunnel current \cite{Has,Wo}. 

In our approach, the filling factors of final states do not enter the
equation for the current Eq. (\ref{current5'}), and the existence 
of SGS is therefore not
obvious, although the side band currents  Eq. (\ref{Kn}) successively diminish
with increasing side band index. However, 
attribution of the nonequilibrium tunnel current in
biased junctions to the current of real excitations is a general
physical argument which should be automatically met in any correct theory.
In fact, and this also follows
from the discussion of the previous section, the true tunnel current is
hidden in Eq. (\ref{current5'}): it results from partial cancellation  
of large contributions of different scattering modes.
The cancellation is nontrivial because of mixture of currents of
different side bands, the odd side
bands containing information about the currents of the even side bands and vice
versa. This means that a finite perturbation expansion of Eq. (\ref{current5'})
is not satisfactory and will not adequately correspond to the perturbative
structure of the true tunnel current. To reveal such a structure one must
rearrange the series in  Eq. (\ref{current5'}). 

To this end we consider a general term $K_n,\;n>0$ in Eq. (\ref{current5'}). 
It follows immediately from the explicit form of the normal and Andreev
transmission coefficients, Eqs. (\ref{fn}) and (\ref{bn}), that the leading term
with respect to $\lambda$ in $K_n$ is proportional 
to a factor $(1-e^{2\Gamma_n})$ which is equal
to zero if $|E_n|<\Delta$. Having made this observation we express the quantity
$K_n$ in the following form:
\begin{mathletters}
\begin{equation}
K_n=
{2\over R}\lambda^n\theta(E_n^2-\Delta^2)e^{-\gamma_n}\sinh\gamma_n|f_n|^2
+4\lambda^{n+1} e^{-2\Gamma_n}\Big|{f_n\over Z_{n+1}}\Big|^2 F_{n+1},\\
\label{1stepa}
\end{equation}
\begin{equation}
\displaystyle
F_{n+1}=|Z_{n+1}|^2+\mbox{Re}\Big({e^{-\gamma_{n+1}}Z_{n+1}^\ast \over
\sinh\gamma_{n+1}}\Big) - \lambda\Big|{e^{-\gamma_{n+1}}\over
\sinh\gamma_{n+1}}\Big|^2.
\label{1stepb}
\end{equation}
\end{mathletters}
In Eq. (\ref{1stepa}) the first term represents the main contribution of the
$n$-th side band to the current: it is proportional to the probability of
normal scattering to the $n$-th side band and it does not contain the
contribution of the side band states lying inside the gap $|E_n|<\Delta$.
Using the recurrence relation (\ref{Zn}) and remembering that $\lambda=D/4R$, 
after some algebra the function $F_n$ in Eq. (\ref{1stepb}) becomes
\begin{mathletters}
\label{Fn}
\begin{equation}
\displaystyle
F_n=
{1\over R}\theta(E_n^2-\Delta^2){1\over \tanh\gamma_n}-
\lambda \Big|{e^{\gamma_n}\over \sinh\gamma_nZ_{n+1}}\Big|^2 G_{n+1},\\
\end{equation}
\begin{equation}
G_{n+1}=|Z_{n+1}|^2-\mbox{Re}\Big({e^{\gamma_{n+1}}Z_{n+1}^\ast \over
\sinh\gamma_{n+1}}\Big) - \lambda\Big|{e^{\gamma_{n+1}}\over
\sinh\gamma_{n+1}}\Big|^2.
\end{equation}
\end{mathletters}
Substituting Eq. (\ref{Fn}) into  Eq. (\ref{1stepa}), we find that the second
term in the equation for $K_n$, 
proportional to $\lambda^{n+1}$, has analytical structure similar to 
the first term in the same equation, proportional to  $\lambda^{n}$, namely,  
it consists of the probability of
normal scattering to the $(n+1)$-th side band [cf. Eq. (\ref{defS})] 
and it does not include
the contribution of the side band states lying inside the gap 
$|E_{n+1}|<\Delta$.
This allows us to associate this term with the effective contribution of the
nearest {\em even} side band.

A similar transformation of the function $G_{n+1}$ in Eq. 
(\ref{Fn}) yields the recurrence:
\begin{equation}
G_{n+1}=
-{1\over R}\theta(E_{n+1}^2-\Delta^2){1\over \tanh\gamma_{n+1}}
-\lambda \Big|{e^{-\gamma_{n+1}}\over \sinh\gamma_{n+1}Z_{n+2}}\Big|^2 F_{n+2}
\label{Gn}
\end{equation}
Combination of Eqs. (\ref{1stepa})-(\ref{Gn}) shows that the next term of
the current $K_n$, proportional to $\lambda^{n+2}$, has the same
analytical structure as the leading term in the current $K_{n+2}$ of the
next odd side band, and therefore it can be regarded as a 
renormalization of that current.

Continuing this procedure by systematic use of the recurrence Eqs.  (\ref{Fn}),
(\ref{Gn}) we obtain the following expansion for the current $K_n$ in 
Eq. (\ref{Kn}):
\begin{equation}
K_n=
{2\lambda^n\over R}\theta(E_n^2-\Delta^2)Q_n
+{4\lambda^{n+1}\over R}\theta(E_{n+1}^2-\Delta^2)e^{-\Gamma_{n}}\cosh\Gamma_{n}
Q_{n+1}
\label{expKn}
\end{equation}
$$
+{4\lambda^{n+2}\over
R}\theta(E_{n+2}^2-\Delta^2)e^{-\Gamma_{n}+2\Gamma_{n+1}}
\cosh\Gamma_{n}Q_{n+2}
$$
$$
+{4\lambda^{n+3}\over R}\theta(E_{n+3}^2-\Delta^2)e^{-\Gamma_{n}+2\Gamma_{n+1}
-2\Gamma_{n+2}}
\cosh\Gamma_{n}Q_{n+3}+...
$$
where we have introduced the quantity $Q_n$ defined for all $n$
as
\begin{equation}
Q_n={e^{\gamma_0} \over \cosh\gamma_0|Z_0|^2}\sinh\gamma_n\cosh\Gamma_n
\prod_{l=1}^{n}{1 \over |Z_l\sinh\gamma_l|^2}
\label{Qn}
\end{equation}

One can derive similar expansions for the currents $\bar K_n$ as well as
for the currents of the side bands with negative $n<0$.
Expanding each term of the series in Eq. (\ref{current5'}) using 
Eq. (\ref{expKn}) and collecting terms with the
same factor $\lambda^n$, we can finally express the series in
the form
\begin{equation}
\sum_{odd}(K_{n}-\bar K_{n})=
\sum_{n\neq 0}(\tilde K_{n}-\tilde{\bar K}_{n}).
\label{newsum}
\end{equation}
The last summation is done over all odd {\em and even} integer $n$, 
and the renormalized coefficients have the form 
\begin{mathletters}
\label{newK}
\begin{equation}
\tilde K_n=\lambda^{n}\theta(E_n^2-\Delta^2)(4Q_n/R)
\left[ (1/2) +\cosh\Gamma_{n-2}e^{-\Gamma_{n-2}+2\Gamma_{n-1}}
\right.
\end{equation}
$$
\left.
+\cosh\Gamma_{n-4}e^{-\Gamma_{n-4}+2\Gamma_{n-3}-2\Gamma_{n-2}+
2\Gamma_{n-1}}+...
+\cosh\Gamma_{1}e^{-\Gamma_{1}+2\Gamma_{2}-2\Gamma_{3}+...+
2\Gamma_{n-1}} \right]
$$
for odd $n>0$, and 
\begin{equation}
\tilde K_n=\lambda^{n}\theta(E_n^2-\Delta^2)(4Q_n/R)
\left[
\cosh\Gamma_{n-1}e^{-\Gamma_{n-1}}
+\cosh\Gamma_{n-3}e^{-\Gamma_{n-3}+2\Gamma_{n-2}-2\Gamma_{n-1}}
 +...\right.\\
\end{equation}
$$
\left. +\cosh\Gamma_{1}e^{-\Gamma_{1}+2\Gamma_{2}-2\Gamma_{3}+...-
2\Gamma_{n-1}} \right]
$$
\end{mathletters}
for even $n>0$. 

The representation of Eqs. (\ref{newsum}),(\ref{newK}) is exact. One can regard a
general term of the series as an effective renormalized current of the $n$-th
side band. In fact, this effective current consists of the contributions of 
all side bands with odd indices smaller then $n$. An important feature of this
representation is the presence of the $\theta$-function in the general term, 
which allows us to separate out in Eq. (\ref{current5'}) the part of the 
current which is obviously responsible for the SGS, 
\begin{equation}
I_{SGS}=\sum_{n=1}^{\infty}{e\over\pi}\int^{neV-\Delta}_{\Delta}
dE\,{E\over\xi}\tanh {E\over 2T}(\tilde K_n-\tilde{\overline K}_n)
\label{kern}
\end{equation}
One might expect (cf. Ref. \cite{FTT}) that Eq. (\ref{kern}) represents the
subgap tunnel current at zero temperature and that the remaining part of the 
current of  Eq. (\ref{current5'}),
\begin{equation}
I_{r}=I-I_{SGS}
=\sum_{n=1}^{\infty}{e\over\pi}
\left[
\int_{neV+\Delta}^{\infty} dE\,
{E\over\xi}
\tanh {E\over 2T}(\tilde K_n-\tilde{\overline K}_n)
+\int_{\Delta}^{\infty} dE\,
{E\over\xi}
\tanh {E\over 2T}(\tilde K_{-n}-\tilde{\overline K}_{-n})
\right],
\label{X}
\end{equation}
corresponds to the current of thermal excitations. However, this separation
is not exact: an analysis shows that the current in 
Eq. (\ref{X}) does not vanish completely at $T=0$, but contributes a small 
residual part. An important property of this residual current is that it does
not contain any structureless component but demonstrates behaviour similar
to the current $I_{SGS}$ in Eq. (\ref{kern}), thus resulting in a small 
correction to Eq. (\ref{kern}).

\section{Subharmonic gap structure.}

The explicit analytical expressions (\ref{kern}) and (\ref{X}) provide a basis 
for numerical calculation of the subgap current for small $\lambda$ 
(low transparency) with any desirable accuracy. 
However, they are also convenient for qualitative discussion of the
SGS. In this section we will analyze the SGS at
zero temperature on the basis of Eq. (\ref{kern}). 

The current-voltage characteristic $I_{SGS}(V)$ in Eq. (\ref{kern}) 
has a complex form consisting 
of a sum of renormalized side band currents
$I_n(V, \lambda)$: 
\begin{equation}
I_{SGS}(V,\lambda)=\sum_{n=1}^\infty I_n(V, \lambda),\;\;\;
I_n(V,\lambda)={e\over\pi}\int^{neV-\Delta}_{\Delta}
dE\,{E\over\xi}(\tilde K_n-\tilde{\overline K}_n).
\label{In}
\end{equation}
The partial current-voltage characteristics $I_n(V, \lambda)$
are similar to each other, and it is convenient to analyze them independently. 

According to  Eq. (\ref{In}) the partial
current $I_n$ starts with an onset at the threshold voltage $V_n=2\Delta/en$.
In the limit $\lambda \rightarrow 0$ the onset is infinitely sharp
and the magnitude of the onset is
\begin{equation}
I_n(V_n,\lambda \rightarrow 0)= e\Delta D^n{2n\over 4^{2n-1}}
{n^{2n}\over (n!)^2}.
\end{equation}
The jumps of the current at the thresholds result from the singular denominators
in Eqs. (\ref{fn}), (\ref{fn'}), related to the singular density of
states at the side band energy gap edges, $\sinh\gamma_k=0$. Accumulation of 
these
singularities in the high order scattering amplitudes yields a tremendous 
increase of the partial currents well above the corresponding thresholds 
- this is what 
causes the failure of multiparticle tunneling theory \cite{W,Has,Wo}.
In our theory, the singularities are regularized by the
factors 	
\begin{equation}
P_n=\prod_{k=0}^n |Z_k|^2
\label{Pn}
\end{equation}
in the denominators of the scattering amplitudes Eq. (\ref{fn}). 
These factors are expressed through the continued fractions $Z_n$, Eq.
(\ref{Zn}), which therefore should be calculated  with sufficient
accuracy to preserve the 
singular parts of $Z_n$ which provide regularization of the integrals.

The first order current $I_1$ in Eq. (\ref{In}) corresponds to direct one
particle scattering to the side band $n=1$, Fig 7a. The explicit form of the 
current $I_1$ is
\begin{equation}
I_1={2e \lambda \over \pi R}
\int_{\Delta}^{eV-\Delta} dE\, 
{|E_1|\over \xi\xi_{1}}
\left( 
{E+\xi\over P_{1}}+
{E-\xi\over \overline P_{1}}
\right).
\label{I1}
\end{equation}
In the limit $\lambda\rightarrow 0$ this current coincides with the
quasiparticle current of the tunnel Hamiltonian model \cite{Co,Wer}. 
At finite $\lambda$ the threshold onset of the current at $V=V_1$ is washed
out. To evaluate the width of the onset we truncate the continued fraction 
in $P_1$ assuming $Z_{-1}=Z_2=1$, obtaining
\begin{equation}
P_1\approx |(1+\lambda a^-_0a^-_{-1})(1+\lambda a^-_1a^-_2) + 
\lambda a^+_0a^+_{1}|^2.
\label{P1}
\end{equation}
The function  $\bar P_1$ has a similar form. The regularization effect of the
threshold singularity is
provided by the most singular term $\lambda a^+_0a^+_{1}$ in
Eq. (\ref{P1}). Keeping this term we obtain in the vicinity of the 
threshold, $e(V-V_1)\ll \Delta$, the following result
\begin{equation}
I_1(V)={2e\Delta \lambda\over \pi R}f\left({eV-eV_1\over\Delta\lambda}
\right),\;\;\;
f(z )=
\int_{0}^{\pi} d\theta\, 
{\sin^2\theta\over (\sin\theta+1/z)^2}.
\end{equation}
According to this formula the onset width is $e(V-V_1)\sim\lambda\Delta$.

The second order current $I_2$ corresponds to the creation of a real 
excitation during quasiparticle backscattering into the side band $n=2$, 
Fig. 7b, and appears as the current of transmitted states of the side band 
$n=1$ (cf. the excess current in Sec. VII). In the vicinity of the 
threshold, $V_2<V<V_1$, this current exists only in the form of currents 
through the bound states and therefore it is completely converted 
into supercurrent far away from the junction. At larger voltages, $V>V_1$, the 
side band $n=1$ extends outside the energy gap (see Fig. 8a), which also makes 
the current $I_2$ partially consist of contributions from extended states.  
The explicit expression for the second order current is 
\begin{equation}
I_2={4e \Delta^3 \lambda^2 \over \pi R}
\int_{\Delta}^{2eV-\Delta} dE\, 
{|E_2|\over\xi\xi_2|\xi_1|^2} 
\cosh\Gamma_1
\left( 
{e^{-\gamma_0+\Gamma_1}\over P_{2}}+
{e^{\gamma_0-\Gamma_1}\over \overline P_{2}}
\right).
\label{I2}
\end{equation}
Omitting the $\lambda$-dependence of $P_2$ in  Eq. (\ref{I2}), one gets 
the two-particle tunnel current of Schrieffer and Wilkins \cite{SW,W}.
To keep the singular terms in $P_2$ one has to truncate the continued
fractions in Eq. (\ref{Zn}) assuming $Z_{-1}=Z_3=1$, which yields
\begin{equation}
P_2\approx
|(1+\lambda a^-_{-1}a^-_0)
(1+\lambda a^-_{1}a^-_2)
+\lambda a^+_{0}a^+_1
(1+\lambda a^+_{2}a^+_3)|^2.
\label{P2}
\end{equation}
The threshold singularity results from the small product $\xi\xi_2$ in
denominator of Eq. (\ref{I2}). However,
there are no singular terms  in Eq. (\ref{P2}) proportional to $a_0a_2$ among 
the terms linear in $\lambda$. Such terms are 
quadratic in $\lambda$ and they provide, along with the terms 
$\lambda a_0$ and $\lambda a_2$, the width of the onset:
$e(V-V_2)\sim\lambda^2\Delta$. This onset is sharper than the onset of
the current $I_1$. 

The threshold singularity in the current $I_2$ is typical
for all higher order currents $n>1$.

The appearance of the first side band outside the energy gap at
$V=V_1$ is manifested through a spike in the current $I_2$. 
Indeed, if $V\approx V_1$, the nodes of $\xi_1$ overlap
the nodes of $\xi$ and  $\xi_2$ at the lower ($E=\Delta$) and  at the
upper ($E=3\Delta$) limits of integration in Eq. (\ref{I2}) respectively (see
Fig. 8a).
This singularity yields an increase of the current $I_2$ when the voltage
approaches $V_1$,
$$I_2\sim e\Delta\lambda^2 \sqrt{{\Delta\over e(V_1-V)}}.$$ 
Regularization of the integral, which is provided  by the singular  
terms $\lambda a_1a_0$ and $\lambda a_1a_2$ in Eq. (\ref{P2}) at the lower 
the and upper integration limits respectively, yields
$$
{I_2(V_1)\over I_2(V_2)}\sim {1\over\sqrt{\lambda}}.
$$
Further analysis shows that maximal magnitude of the current is achieved
slightly above $V=V_1$, after which the current rapidly decreases (see Fig. 9).
At voltages $V>V_1$ the singular point $\xi_1=0$ remains 
within the integration region, which gives rise to enhancement of the
magnitude of the current by a logarithmic factor in comparison with the
current magnitude near the threshold $V_2$, 
\begin{equation}
I_2(V>V_1)\sim{e\Delta \lambda^2\ln\lambda \over R}.
\end{equation}
At large voltage $V\gg V_1$ the current $I_2$ forms
the excess current Eq. (\ref{excess}). It is interesting to note, that
in this limit the logarithmic factor is
compensated for by the current $I_r$, Eq. (\ref{X})  which yields
the $\lambda^2$-dependence of the excess current. 

The third order current $I_3$ at voltages close to the threshold $V_3$ 
results from the combination of one-particle tunneling into the side band $n=3$ 
and excitation of
the transmitted Andreev bound states of the side band $n=1$, Fig. 7c. 
The probabilities of these two processes are related as 1:2  at 
threshold, Eq. (\ref{newK}). 
Successive emergence of the bound states of the side bands $n=1$ and 
$n=2$ outside the energy gap at $V=V_2$ and $V=V_1$, Fig. 8b,
gives rise to the current peaks. The current $I_3$ has the explicit form 
\begin{equation}
I_3={2e\Delta^5\lambda^3 \over \pi R}
\int_{\Delta}^{3eV-\Delta} dE\, 
{|E_3|\over \xi\xi_{3}|\xi_1\xi_2|^2}
\left[ 
{e^{\gamma_0}(1+2\cosh\Gamma_1e^{-\Gamma_1+2\Gamma_2})\over P_{3}}
+{e^{-\gamma_0}(1+2\cosh\Gamma_1e^{\Gamma_1-2\Gamma_2})\over \overline P_{3}}
\right]
\end{equation}
with the regularization factor 
\begin{equation}
P_3\approx
|(1+\lambda a^-_{-1}a^-_0)
(1+\lambda a^-_{1}a^-_2)
(1+\lambda a^-_{3}a^-_4)
+\lambda a^+_{0}a^+_1
(1+\lambda a^+_{2}a^+_3)|^2.
\label{P3}
\end{equation}
The current peak at $V=V_2$ results from the overlap of 
nodes of $\xi$ and $\xi_2$ at $E=\Delta$ and nodes of 
$\xi_1$ and $\xi_3$ at $E=2\Delta$, similarly to the peak of the current 
$I_2$. These singularities yield again an
increase of the current inversely proportional to the square root of the 
departure from the
voltage $V_2$: $I_3\sim e\Delta^{3/2}\lambda^3/\sqrt{e(V_2-V)}$. However, since the
factor $P_3$, Eq. (\ref{P3}), contains neither a term $\lambda a_0a_2$
nor a term $\lambda a_1a_3$, regularization of the singularity is provided,
e.g. at $E=\Delta$, by the terms $\lambda a_0$ or $\lambda a_2$, which gives
rise to a more pronounced peak with magnitude
\begin{equation}
{I_{3}(V_2)\over I_3(V_3)}\sim {1\over\lambda}.
\end{equation}
We note, that the magnitude of this peak is comparable with the magnitude
of onset of the current $I_2$.
The second peak at $V=V_1$ results from the overlap of nodes of $\xi_2$ 
and $\xi_3$ at
$E=3\Delta$; this gives rise to increase of the current $ I_3$ near the
voltage $V=V_1$ inversely proportional to the first power of the
distance to this voltage: $I_3\sim \lambda^3\Delta^2/(V_1-V)$. The 
divergence is regularized by the term  $\lambda a_2a_3$ in Eq. (\ref{P3}) 
which results in a peak with magnitude 
\begin{equation}
{I_3(V_1)\over I_3(V_3)}\sim {1\over\lambda}.
\end{equation}
Thus the heights of both peaks of the current
$I_3$ are of the same order in $\lambda$, although the peak at $V\approx
V_1$ is sharper. 

In a similar way, all of the high-order currents result, in the vicinity of 
their thresholds, either from Andreev bound state currents (even $n$) or
from a combination of Andreev bound state currents and the current of a single
real excitation (odd $n$). The number of excited Andreev states is
correspondingly $n/2$ or $(n-1)/2$.
Singularities similar to the singularity of the current $I_3$ at the
voltage $V\approx V_1$ exist in all high
order currents, where they cause even more pronounced current peaks 
because of
the absence of terms  $\lambda a_ka_{k+1}$ in the corresponding smearing 
functions $P_n$. Due to this property, the magnitudes of such peaks exceed 
the threshold magnitude of the corresponding current 
by two orders of $\lambda$: $(I_{n})_{\mbox{max}}\sim
e\Delta\lambda^{n-2}/R$.

The above discussion reveals the current peaks to be essential features of 
the SGS
of tunnel current in addition to the current onsets, Fig. 9 (these peaks are 
seen also in the numerical results of Refs.\cite{Ave,Arn2}). It allows us to 
establish a general classification of singularities causing peaks in partial 
currents $I_n$. They result from the overlap of singularities of the side band
density of states. It is easy to see that the singularities of only 
two side bands can overlap, the condition of the overlap for $m$-th
and $k$-th side bands having the form
\begin{equation}
E-keV=\Delta, \;\;\;
E-meV=-\Delta.
\end{equation}
This condition is met at voltages $eV=2\Delta/(m-k)$ for all integer $0\leq
k<m\leq n$. The magnitude 
of the current peaks depends on whether the overlapping side bands are
neighbours or not, and whether the side band index is inside or at the edge 
of the interval $(0,n)$.

I. $m-k=1$,  $m=n$ or $k=0$: edge-type singularity, neighbour side bands. 
This type of singularity
forms the peak of the current $I_2$ at the main threshold $V_1$. The
magnitude of the current peak is:
$(I_{2})_{\mbox{max}}\sim e\Delta\sqrt{\lambda}/R$.

II. $m-k>1$, $m=n$ or $k=0$: edge-type singularity, non-neighbour side bands. 
This type of singularity forms
the first peak of each current $I_n$, $n>2$ at voltage $V_{n-1}$.
The magnitude of the current peak is:
$(I_{n})_{\mbox{max}}\sim e\Delta \lambda^{n-1}/R$. 

III. $m-k=1$, $m<n,\; k>0$: internal singularity, neighbour side bands. 
This type of singularity forms
the last peak of each current $I_n$, $n>2$ at voltage $V_1$.
The magnitude of the current peak is:
$(I_{n})_{\mbox{max}}\sim e\Delta\lambda^{n-1}/R$. 

IV. $m-k>1$, $m<n,\; k>0$: internal singularity, non-neighbour side bands. 
This type of singularity forms
all intermediate peaks of each current $I_n$, $n>3$.
The magnitude of the current peaks are:
$(I_{n})_{\mbox{max}}\sim e\Delta\lambda^{n-2}/R$. 

\section{Conclusion}

In this paper we have considered superconductive tunneling as a scattering
problem within the framework of Bogoliubov-de Gennes (BdG) quantum
mechanics. An essential aspect of this problem is that the scatterer 
consists not only of the potential of the tunnel barrier but also of the 
discontinuity of the phase of the order parameter. In equilibrium 
(zero bias, dc Josephson current) the
scattering problem is {\em elastic}. The peculiar feature of the elastic
scattering problem in short junctions, considered here, is that the balance
of currents of the scattering modes is not violated: the supercurrent
flows only through the superconducting bound states (for a more general
discussion, see \cite{WSh}). In the presence of voltage bias the scattering
is {\em inelastic} due to the time dependence of the component of the
scatterer related to the superconducting phase difference. Generally, 
the currents of
all inelastic channels constitute together the components of the tunnel 
current flowing through
the biased junction: the quasiparticle current corresponds to the
incoherent part of the inelastic side band contributions, and the ac
Josephson current corresponds to the interference of the side band
contributions.

There are three distinct components of the quasiparticle
tunnel current at zero temperature: (i) the current of quasiparticles
excited
above the ground state, (ii) the current trough Andreev bound states
converted to supercurrent outside the junction, and (iii)
the imbalance current of the ground state modes.
At large bias voltage, $eV\gg 2\Delta$, the first component corresponds to
the single particle current of the  normal junction, while
the other components cause excess
current.  When voltage is decreased, redistribution of current among the
components gives rise to subharmonic gap structure (SGS) in the form of
current onsets and current peaks.
Within the voltage intervals $2\Delta/n<eV<2\Delta/(n-1)$ with even $n$,
the tunnel current entirely consists of currents through the Andreev bound
states [component (ii); e.g. Fig. 7b], the states of all side bands with
odd 
indices smaller than $n$ contributing to the current. If $n$ is odd, a real
excitation current of the side band $n$ [component (i); e.g. Fig. 7a,c] is
also present in the tunnel
current. Opening of new channels of tunneling of real excitations gives
rise to current structures. Thus, SGS reveals the discrete nature of the
side band spectrum. The structure is the more pronounced
the smaller is the transparency of the junction.

Since each Andreev state provides transfer of one Cooper pair
through the junction for every incident quasiparticle, $n$ particles will
tunnel in the interval $2\Delta/n<eV<2\Delta/(n-1)$. One may be surprised 
by the participation of a large number of bound Andreev states
in current transport at low voltages: this appears to contradict the fact
that subgap current diminishes at zero bias. After all, the probability
of the scattering into side bands does not depend on the bias and is
proportional to powers of $D$. The solution to this paradox results
from increasing compensation of currents between the normal and Andreev
channels in each side band with decreasing voltage, which gives the 
required voltage dependence of the total current.

\section{Acknowledgement}

This work has been supported by the Swedish National Science Research
Council (NFR) and Swedish National Board for Technical and Industrial
Development (NUTEK), and the Swedish Royal Academy of Sciences (KVA).

\appendix
\section{Boundary conditions}

The quasiclassical boundary condition in Eqs. (\ref{match}), (\ref{V}) has been
derived in  Ref.\cite{Sh} using the method of Ref. \cite{Svi}. Here we present
simple arguments which lead to this boundary condition. We consider the more
general case of asymmetric junction, using an asymmetric version of the  
Hamiltonian of  Eq. (\ref{Hamiltonian}) with the same restriction on the 
length of the non-superconducting region: $L\ll\xi_0$. We include a contact
potential difference into the potential $U(x)$ which implies that this
potential may have non-vanishing asymptotic values at infinity: $U(-\infty)\neq
U(\infty)\neq 0$. If the junction has more than one transverse transport mode
we assume that these modes are not mixed.

A one-dimensional quasiclassical wave function of a given transverse channel
in the right electrode has the form [Eq. (\ref{quasiclassic})],
\begin{equation}
\Psi_R(x,t)=\sum_{\beta}{1\over\sqrt{v_R}}
e^{i\beta\int p_Rdx}e^{i\sigma_z\chi_R/2}\psi_R^\beta(x,t),
\label{quasiclassic'}
\end{equation}
with similar expression for the left electrode.
$\psi_R^\beta$ are slowly varying two-component wave functions on the scale 
of $1/p_R$, where $p_R(x)=\sqrt{2m_R(\mu-U_R-E_{\perp R}(x))}$. This equation is
valid over the distance $x\gg 1/p_R$ from the junction, and in the spatial
region $1/p_R\ll x\ll \xi_0$ the functions $\psi_R^\beta$ are almost
constant.

From another point of view, at large distance from the junction, $|x|\gg
1/p_{R,L}$, the function $\Psi$ can be expressed in the form of a 
linear combination of the scattering states at the Fermi level,
\begin{equation}
\Psi=C_1\chi_1 + C_2\chi_2
\label{chi12}
\end{equation}
\begin{mathletters}
\label{chi}
\begin{equation}
\chi_1=\left\{
\begin{array}{lr}
\displaystyle
(1/\sqrt{v_L})[e^{ip_Lx}+re^{-ip_Lx}],& x < 0 \\
\displaystyle
(1/\sqrt{v_R})de^{ip_Rx}, & x > 0 ,
\end{array}
\right.
\end{equation}
\begin{equation}
\chi_2=\left\{
\begin{array}{lr}
\displaystyle
(1/\sqrt{v_R})[e^{-ip_Lx}+\tilde re^{ip_Rx}],& x >0 \\
\displaystyle
(1/\sqrt{v_L})\tilde de^{-ip_Lx}, & x< 0 .\\
\end{array}
\right.
\end{equation}
\end{mathletters}
Comparing  Eqs. (\ref{chi12}), (\ref{chi}) with Eq. (\ref{quasiclassic'}) 
in the region $1/p_F\ll |x|\ll \xi_0$ we have
\begin{equation}
C_1=e^{i\sigma_z\chi_L/2}\psi^+_L, \;\;
C_2=e^{i\sigma_z\chi_R/2}\psi^-_R,\;\;
rC_1+\tilde dC_2=e^{i\sigma_z\chi_L/2}\psi^-_L,
\end{equation}
$$
dC_1+\tilde rC_2=e^{i\sigma_z\chi_R/2}\psi^+_R,
$$
which yields the boundary condition 
\begin{equation}
\pmatrix{\psi^-\cr \psi^+\cr}
=\hat V
\left(\begin{array}{c}
\psi^+\\
\psi^-
\end{array}\right),
\end{equation}
with the matching matrix
\begin{equation}
\hat V=
\left(
\begin{array}{cc}
r & \tilde de^{i\sigma_z\phi/2}\\
de^{-i\sigma_z\phi/2} & \tilde r
\end{array}\right),
\end{equation}
where $\phi=\chi_R(0)-\chi_L(0)$. The matrix $\hat V$ satisfies the
unitarity condition $\hat V\hat V^\dagger=1$, provided by the relations
among the normal electron scattering amplitudes in Eq. (\ref{chi}):
$\tilde r=-\tilde d(r/d)^\ast, \;\;|d|^2=|\tilde d|^2=D, \;\;
|r|^2=|\tilde r|^2=R=1-D$.

\section{Bound state current}

Equation (\ref{dE/dphi}) for the current of a single bound state can be
derived directly \cite{Gor} from the Bogoliubov-de Gennes equation 
Eq. (\ref{BdG0}), (\ref{Hamiltonian}). The derivation is
valid for junctions with an arbitrary non-superconducting region between
superconducting electrodes.
We assume for simplicity that the phase of the order parameter, Eq.
(\ref{Delta}), in the electrodes is constant and equal to $\pm\phi/2$ 
in the right and left electrodes respectively. 
Let $\Psi(\vec r,E)$ to be a normalized wave function of the Andreev bound
state with energy E,
\begin{equation}
\hat H\Psi-E\Psi=0, \;\;\;\Psi(x=\pm\infty)=0.
\label{HE}
\end{equation}
The energy and the wave function of the bound state depend on the phase
difference $\phi$. 
Taking the derivative with respect to $\phi$ of Eq. (\ref{HE}) and
considering a scalar product of the resulting equation with the function 
$\Psi$, we have
\begin{equation}
\int d^3r(\Psi, {d\over d\phi}(\hat H-E) \Psi)=0,
\label{HE'}
\end{equation}
where the brackets denote a scalar product in electron-hole space,
similar to Eq. (\ref{normal}). In this equation the derivative of the
Hamiltonian has the form
\begin{equation}
{d\hat H\over d\phi}= {d\hat \Delta\over d\phi}=
{i\mbox{sign} x\over 2}\sigma_z\hat\Delta,
\label{Dphi}
\end{equation}
in accordance with Eqs. (\ref{Hamiltonian}), (\ref{Delta}). 
Substituting relation (\ref{Dphi}) into
Eq. (\ref{HE'}) and taking into account that the function $\Psi$ is
normalized, we get
\begin{equation}
{dE\over d\phi}=\int d^3r(\Psi, {d\hat\Delta\over d\phi} \Psi).
\label{Ephi}
\end{equation}

The continuity equation for the charge current,
\begin{equation}
I(x,E)={e\over 2m}(\hat p_x-\hat p'_x)\int d^2r_\perp(\Psi(x'),\Psi(x))
_{x=x'},
\end{equation}
in accordance with Eq. (\ref{HE}) has the form
\begin{equation}
i{d\over dx}I(x,E)=e\int d^2r_\perp(\Psi(x),[\sigma_z,\hat\Delta]\Psi(x)).
\label{Icom}
\end{equation}
Substituting relation (\ref{Dphi}) into Eq. (\ref{Icom}) and performing
integration of this equation over the whole $x$ axis we get
\begin{equation}
I(0,E)=2e\int d^3r(\Psi,{d\hat\Delta\over d\phi}\Psi).
\label{IDphi}
\end{equation}
In equation (\ref{IDphi} the current at infinity drops out 
because of decay of the bound state wave function, $I(\pm\infty)=0$. The
magnitude $I(0)$ is formally taken in the middle of the junction; however, 
the current has the same magnitude along the whole non-superconducting 
region according to the conservation equation (\ref{Icom}). 
Comparison of Eqs. (\ref{Ephi}) and (\ref{IDphi}) finally yields
\begin{equation}
I(E)=2e{dE\over d\phi}.
\end{equation}

\begin{figure}
\label{fig1}
\caption{SIS tunnel constriction.}
\end{figure}

\begin{figure}
\caption{Quasiparticle spectrum and position of the incoming states:
1(3) - hole (electron)-like quasiparticle incident from left; 2(4) -
hole (electron)-like quasiparticle incident from right.}
\end{figure}

\begin{figure}
\caption{Spatial configuration of the edges of the superconducting energy
bands in a long constriction:
$E_{min}, E_{max}=\pm\Delta+p_s(x)v$. A potential well appears in upper
(lower)
band for electrons moving in a direction opposite to (along) the
supercurrent.}
\end{figure}

\begin{figure}
\caption{Scattering diagram of voltage biased superconducting tunnel
junctions.
Solid (dotted) arrows indicate scattering in
the normal (Andreev) channel. Filled triangles indicate superconducting
bound states. Transmission (reflection) occurs into side bands with odd
(even) indices.}
\end{figure}

\begin{figure}
\caption{Scattering diagrams of voltage biased normal tunnel junctions.
a) scattering of normal holes with spectrum $E_h=\mu-(p^2/2m)$;
represents an elementary fragment of the diagram in Fig.
4 for $j=1$. b) scattering of the normal electrons with spectrum
$E_e=p^2/2m$. c) conventional diagram of elastic electron scattering in
biased tunnel junctions; the local chemical potentials in the electrodes
are
then shifted by $eV$}
\end{figure}

\begin{figure}
\caption{Three kinds of processes contributing to the tunnel current at
large bias $eV\gg\Delta$: a) creation of a real excitation across the gap
by forward scattering; b) excitation of the Andreev bound state due to
creation of
a real excitation via backward scattering (dashed arrow); c) imbalance of
ground state modes due to creation of a real excitation via backward
scattering.
Excess current is caused by processes b) and c).}
\end{figure}

\begin{figure}
\caption{Scattering processes contributing to the subgap current.
a) single
particle scattering into side band $n=1$ gives the main contribution at
$eV>2\Delta$. b)  excitation of the Andreev bound state ($n=1$) due to
backward scattering into side band $n=2$ gives the main contribution at
$eV>\Delta$. c) single
particle scattering into side band $n=3$ and simultaneous excitation of the
Andreev bound state in side band $n=1$ gives the main contribution at
$eV>2\Delta/3$.}
\end{figure}

\begin{figure}
\caption{a) Density of states $\nu (E)=|E_n/\xi_n|$ of the side bands
$E_0,E_1,E_2$ at applied voltage $V>V_1$ (right), position of
singularities
of side band density of states as function of applied voltage for current
$I_2$
(left), $1^\pm:E_1=\pm\Delta$, $2^-: E_2=-\Delta$.
b) Density of states  of the side bands $E_0,E_1,E_2,E_3$ at applied
voltage
$V_2<V<V_1$ (right), position of singularities
of side band density of states as function of applied voltage for current
$I_3$
(left), $1^\pm: E_1=\pm\Delta$, $2^\pm: E_2=\pm\Delta$, $3^-:
E_3=-\Delta$.}
\end{figure}

\begin{figure}
\caption{Schematic picture of  partial I$_n$-V characteristics.}
\end{figure}


\begin{thebibliography}{999}
\bibitem{note1}
The contribution from the continuum does not vanish if there is imbalance
of
populations of electron-like and hole-like branches of the quasiparticle
spectrum in one of the electrodes.
\end{thebibliography}
\end{document}